\pdfoutput=1
\documentclass[11pt]{article}
\usepackage{graphicx}
\usepackage{subfigure,amssymb,amsmath}
\usepackage{cite,color,xcolor,url,cancel,ulem,float}
\usepackage{bm}
\usepackage{float}
\usepackage{jheppub}
\newcommand{\cred}[1]{{\color{red}#1}}
\newcommand{\cblue}[1]{{\color{blue}#1}}

\newcommand{\cmag}[1]{{\color{magenta}#1}}

\newcommand{\beq}{\begin{equation}}
\newcommand{\eeq}{\end{equation}}
\newcommand{\bea}{\begin{eqnarray}}
\newcommand{\eea}{\end{eqnarray}}
\def\alphas{\alpha_{_S}}

\def\tanb{\tan\beta}
\def\yb{y_b}
\def\ytau{y_\tau}
\def\ybtau{y_{b,\tau}}
\def\abprime{A_b^\prime}

\def\atauprime{A_\tau^\prime}
\def\abtauprime{A_{b,\tau}^\prime}

\def\mbtau{m_{b,\tau}}
\def\muprime{\mu^\prime}

\def\hsm{h_\mathrm{SM}}
\def\sQ3{\widetilde{Q}_3}
\def\sU3{\widetilde{U}_3}
\def\sD3{\widetilde{D}_3}

\def\sbleft{\tilde{b}_L}
\def\sbright{\tilde{b}_R}

\def\sbone{\tilde{b}_1}
\def\sbtwo{\tilde{b}_2}
\def\sbottom{\tilde{b}}

\def\sbones{{\tilde{b}}^*_1}
\def\sbtwos{{\tilde{b}}^*_2}
\def\ntrli{\chi_i^0}

\def\charpm{\chi^\pm}

\def\mone{M_1}
\def\mtwo{M_2}
\def\mthree{M_3}
\def\mgluino{m_{\tilde{g}}}

\def\mhsm{m_{h_{\mathrm{SM}}}}

\def\msbleft{m_{\tilde{b}_L}}
\def\msbright{m_{\tilde{b}_R}}
\def\mstone{m_{\tilde{t}_1}}
\def\msttwo{m_{\tilde{t}_2}}
\def\msb{m_{\tilde{b}}}
\def\msbone{m_{\tilde{b}_1}}
\def\msbtwo{m_{\tilde{b}_2}}
\def\msbonetwo{m_{\tilde{b}_{1,2}}}
\def\mstaul{m_{\tilde{\tau}_L}}
\def\mstaur{m_{\tilde{\tau}_R}}

\def\bbbar{b\bar{b}}
\def\ttbar{t\bar{t}}
\def\tautaubar{\tau\bar{\tau}}
\usepackage[numbers,sort&compress]{natbib}
\title{Associated production of heavy Higgs bosons with a $\bbbar$ pair in the Nonholomorphic MSSM and LHC searches} 
\author[a]{Utpal Chattopadhyay,}
\affiliation[a]{School of Physical Sciences, Indian Association for the Cultivation of Science,\\  
	2A \& B Raja S.C. Mullick Road, Jadavpur, 
	Kolkata 700 032, India}
\author[b]{AseshKrishna Datta,}
\affiliation[b]{Harish-Chandra Research Institute, A CI of Homi Bhabha National Institute, Chhatnag Road,
	Jhunsi, Prayagraj (Allahabad) 211019, India}
\author[c]{Samadrita Mukherjee,}
\affiliation[c]{Department of Theoretical Physics, Tata Institute of Fundamental Research, 1 Homi Bhabha Road, Colaba, Mumbai 400005, India}
\author[d]{Abhaya Kumar Swain}
\affiliation[d]{School of Physics and Institute for Collider Particle Physics, University of the
Witwatersrand, Johannesburg, Wits 2050, South Africa}
\emailAdd{tpuc@iacs.res.in}
\emailAdd{asesh@hri.res.in}
\emailAdd{samadrita.mukherjee@tifr.res.in}
\emailAdd{abhaya.kumar.swain@cern.ch}

\abstract{In the NonHolomorphic Supersymmetric Standard Model (NHSSM), the 
Yukawa couplings of the bottom quark ($y_b$) and the tau lepton ($y_{\tau}$) might 
receive substantial supersymmetric (SUSY) radiative corrections which have 
prominent dependencies on the NHSSM-specific trilinear soft parameters,
$A_b^\prime$ and $A_{\tau}^\prime$, respectively, in addition to their well-known 
dependence on $\tan\beta$ as is already present in the Minimal SUSY Standard Model 
(MSSM). We study to what extent these could affect the production cross 
sections of the heavy Higgs bosons ($H$ and $A$) in association with a pair of 
$b$-quarks and their decay branching ratios, in particular, to a $\tau\bar{\tau}$ 
pair and compare them with those obtained in the MSSM. Requiring compliance with the 
recently observed upper bounds on the product of their total cross section and 
the branching ratio to $\tau\bar{\tau}$ at the 13 TeV run of the Large Hadron 
Collider (LHC), with data worth 139 fb$^{-1}$,  results in an altered exclusion region in the customary $m_A-\tan\beta$ plane in the framework of the NHSSM when compared to what is derived by the LHC experiments within an MSSM setup. Such alterations are estimated to be pronounced only for large $\tan\beta$ ($\geq 40$) and for large, negative $A_b^\prime$ when one finds a reinforced exclusion of the $m_A-\tan\beta$ plane with an excluded $m_A$ value larger by
$\approx 200$ GeV, compared to the MSSM case, at $\tan\beta=60$. On the other hand, the maximum relaxation in $m_A$, for a similarly large but positive $A_b^\prime$, barely exceeds $\approx 100$ GeV as a result of complementary variations in production cross sections and decay branching fractions of the heavy, neutral Higgs bosons.
}


\begin{document}
%
\maketitle
%
\section{Introduction}
Ever since its discovery at the Large Hadron Collider (LHC) the properties of 
`the' Higgs boson are being probed in minute details thanks to the increasing 
volume of the accumulated data. While its mass, which is not predicted by the 
theory, is being estimated with an ever-increasing precision, various couplings of 
the Higgs boson are being measured with even greater accuracy. It is now known 
that whereas its couplings to the gauge bosons agree with their predictions from 
the Standard Model (SM) at the $\sim 5$\% level, the same with the fermions from 
the third generation, to date, could attain compatibility with the SM only at 
the level of 15-20\% \cite{Aad:2019mbh}.

The latter set of couplings are the Yukawa couplings ($y_b$, $y_t$ and $y_\tau$) 
that are proportional to the respective fermion (bottom, top and tau) masses. 
These are rather special in the sense that, when compared to the minuscule ones 
for the light fermions from the first two generations, these are much larger. In 
addition, the magnitudes of such couplings could also be sensitive to scenarios 
beyond the SM, the implications of which could well transcend the exclusive domain 
of particle physics
by possibly influencing the physics of the early Universe. Direct experimental 
probes to such couplings thus have to be of paramount importance.

It is appreciated that any such fundamental coupling of a given excitation, when 
could appear both in its production and decay, should be determined by studying 
them in tandem to reduce the involved uncertainties in its estimation. In the 
current context of the SM(-like) Higgs boson ($\hsm$), such a combined approach 
is inherently unavoidable given its production cross section at the LHC in any 
relevant mode and its decays (branching fractions) could get to be largely 
interlaced and hence the couplings are not independently extricable
\cite{Wiesemann:2014ioa}. As for the Yukawa couplings, each of $y_t$ and $y_b$ 
might play a role in both production and decay of the Higgs boson. 
However, as pointed out in reference \cite{Wiesemann:2014ioa}, the possibility 
of an efficient extraction of $y_b$ in the decay $\hsm \to \bbbar$ is marred by 
several issues even though this has the largest branching fraction. A priori, a 
better sensitivity to $y_b$ has been expected in the associated $\hsm\bbbar$ 
production mode notwithstanding the large irreducible background that it 
attracts. This is since such a mode is much less encumbered with theoretical 
assumptions when compared to working with the decay branching fraction BR[$\hsm \to 
\bbbar$] which depends on to what extent $\hsm$ decays to other modes and can be
highly model-dependent. However, while the LHC experiments are yet to take a 
dedicated look into the $\hsm \bbbar$ production process (which has a cross section 
similar to that of $\hsm\ttbar$ production that has already been studied by the 
experiments), a recent study \cite{Pagani:2020rsg} has cast a shadow on its 
prospects. 

Under the circumstances, studying productions of various Higgs bosons, possibly heavier 
than $\hsm$, in association 
with a $\bbbar$-pair in scenarios with extended Higgs sector assumes a 
heightened significance. Extensions of the SM endowed with two Higgs doublets, of 
which the Minimal Supersymmetric extension of the SM (MSSM) is a popular example, 
are in immediate reference here. Such production modes could be the direct probes 
to such scenarios. This is since in these scenarios $y_b$ could get 
significantly enhanced for large values of $\tanb \, (={v_u \over v_d})$, the 
ratio of the vacuum expectation values ({\it vev's}), $v_u$ and $v_d$, of the 
neutral components of the two $SU(2)$ Higgs doublets with opposite hypercharges, $h_u$ and $h_d$. The former gives tree-level masses to the up-type quarks while the latter is responsible for the same for the down-type quarks and the charged leptons of the scenario. In addition, the interaction 
vertices of Higgs bosons and a $\bbbar$ pair have an additional $1/\cos\beta$ 
dependence which lead to their enhanced interaction strengths for larger values of $\tanb$. Even then, it has been observed that unless $y_b$ gets significantly 
enhanced in such new physics scenarios, a production process like $\hsm 
\bbbar$ is unlikely to be useful for the purpose. It has been recently 
pointed out \cite{Pagani:2020rsg} that such an enhancement in $y_b$ is strongly 
disfavored by the current measurements of the $\hsm \to \bbbar$ decay. This is 
consistent with the implications of the recent bounds on the $m_A$-$\tanb$ plane 
\cite{Aad:2020zxo,Aaboud:2017sjh,Sirunyan:2018zut}, $m_A$ being the mass of the $CP$-odd Higgs boson of the MSSM scenario, which indicates that the $\hsm \bbbar$ coupling may not be reinforced enough even when $y_b$ is significantly enhanced because of a suppressed mixing between the two $CP$-even Higgs states of the MSSM.

In this backdrop, a renewed impetus for carrying out such studies can now 
be drawn from the recent discourses of an otherwise MSSM scenario but for the 
presence of additional `nonholomorphic' soft terms in its Lagrangian which goes by 
the name of Nonholomorphic Supersymmetric Standard Model (NHSSM)\cite{Girardello:1981wz, Martin:1999hc, Ross:2016pml, Ross:2017kjc, Chattopadhyay:2017qvh, Chakraborty:2019wav, Haber:2007dj, Bagger:1993ji, Ellwanger:1983mg, Jack:1999ud}. 
It is further known that the presence of generic nonholomorphic terms like $A_f' \phi^* \phi$ could have important implications \cite{Chattopadhyay:2016ivr}. In particular, the nonholomorphic trilinear soft term from the bottom quark sector involving
$\abprime$ and $\atauprime$ might play important roles in the phenomenology of the NHSSM  as these could radiatively alter the magnitudes of $\yb$ and $\ytau$, respectively \cite{Chattopadhyay:2018tqv}, over and above what is induced by $\tanb$ in the MSSM \cite{Hall:1993gn, Hempfling:1993kv, Carena:1994bv, Pierce:1996zz, Logan:2000cz, Antusch:2008tf} (see references
\cite{Carena:2002es} and \cite{Djouadi:2005gj} for an overview in the MSSM case). 
Even then, the couplings $\hsm\bbbar/\tautaubar$ remain suppressed because of the reason mentioned in the previous paragraph. In a complementary way, the heavier Higgs bosons of the scenario are
well-poised to receive the maximal benefit of the suppressed Higgs mixing angle and hence their productions 
in association with a $\bbbar$ pair might emerge to be more potential probes to 
$y_b$. However, the LHC bound mentioned earlier that is
obtained assuming the MSSM framework is expected to be sensitive to the 
parameter(s) (like $\abprime$) of the NHSSM. Hence a recast of the bound in the 
NHSSM scenario would be necessary before drawing a conclusion.  

Radiatively corrected $y_b$ and $\ytau$ in the NHSSM scenario are expected to affect the constraints derived in 
the MSSM framework by the LHC experiments on the heavy Higgs boson sector in the form 
of an exclusion in the $m_A-\tanb$ plane. This is since these studies search for 
heavier Higgs bosons in their productions in association with $b$-quark(s), i.e., $pp 
\to \bbbar H/A$ \cite{Aad:2020zxo,Aaboud:2017sjh,Sirunyan:2018zut}, $btH^\pm$ 
\cite{ATLAS:2018gfm, ATLAS:2021upq}, and their subsequent decays to $\tau$-leptons (and to top quarks, in the case of $H^\pm$), i.e., $H,A \to \tautaubar, \, H^\pm \to \tau \nu_\tau$ \cite{ATLAS:2018gfm}, $tb$ \cite{ATLAS:2021upq}. However, an understanding of the extents to which $\yb$ and $\ytau$ could actually influence the proceedings requires a somewhat detailed analysis which we take up duly in this work.
As for the $H^\pm$ states, their searches at the LHC, as yet, could only set much weaker exclusions on the $m_A-\tanb$ plane when compared to those from the searches of the
$H/A$, except for rather low values of $\tanb$ ($\lesssim 2$) which are anyway not so relevant for the phenomenology of the latter states that we are interested in. Hence we do not discuss the related phenomenology of the $H^\pm$ states separately in 
the present work.

Production processes of Higgs bosons in association with bottom squarks ($\sbottom$) also involve $\yb$ and hence are, a priori, in context. However, we find that given the current ATLAS lower bounds on the mass of the lighter sbottom ($\sbone$)
\cite{ATLAS:2021yij}, the maximum $\sigma_{(pp \to \sbone \sbones H)}$ obtainable over the NHSSM parameter space for a relatively light `$H$' state is smaller by quite a few orders when compared to $\sigma_{(pp \to \bbbar H)}$ for the same $m_H$. For our
purposes, we add to the latter  the contribution $\sigma_{(pp \to \bbbar A)}$ where $m_A \simeq m_H$. In contrast, note that $\sigma_{(pp \to \sbone \sbones A)}$ vanishes since the vertex $A \sbone \sbone$ is 
absent \cite{Dedes:1998yt} and $\sigma_{(pp \to \sbone \sbtwos A+ \mathrm{c.c.})}$ is 
further suppressed due to a more massive $\sbtwo$ in the final state. Thus, the combined rate for $H,A$ production in association with a pair of sbottoms is way too small to be sensitive to the contemporary runs of the 
LHC. Hence we do not consider those processes further. We are then set free to consider much larger soft SUSY breaking masses for the sbottoms (as well as, for the staus),
$m_{\tilde{b}, \tilde{\tau}}$.
This, in  turn, allows us to study a wider range of values for the phenomenologically 
interesting NHSSM parameters like $\abtauprime \, (\lesssim m_{\tilde{b},\tilde{\tau}})$ without running the risk of 
encountering a tachyonic spectrum or a charge and color breaking (CCB) minimum in the scalar potential of the theory or jeopardizing the stability of the electroweak vacuum \cite{Beuria:2017gtf}. Henceforth, we refer to these issues collectively as the CCB problem.

The paper is organized as follows. In section 2 we outline the theoretical setup of 
the NHSSM scenario under study with an emphasis on how the same could provide 
additional radiative corrections to running $y_{b,\tau}$ (in the presence of a 
nonvanishing $\abtauprime$) over and above the ones found in the MSSM
($\tanb$-enhancement)) and how these new effects modify the couplings of the neutral 
Higgs bosons to a $b$-quark or a $\tau$-lepton pair. Section 3 is devoted to a 
detailed study of decays of the heavier neutral Higgs bosons, `$H$' and `$A$', to a
$\bbbar$ and a $\tautaubar$ pair (the only accessible final states, following 
relevant experimental studies) for varying $\tanb$ and $\abprime$. This is followed 
by a study of similar variations of the combined cross section for the associated
$\bbbar \Phi$ ($\Phi \equiv H, A$) productions at the 13 TeV LHC where we discuss 
briefly some involved theoretical considerations that go into obtaining a
state-of-the-art, precise estimate of these cross sections. The variation of the 
quantity $\sigma_{_{(pp \to \bbbar H,A)}} \times \mathrm{BR}[H,A \to \tautaubar]$ is 
then studied in detail over the relevant parameter space of the NHSSM and then 
confronted by the experimental bound on the same. The results are then recast on the 
$m_A-\tanb$ plane with $\abtauprime$ as parameters to find the ways in which the 
exclusion contour obtained by the LHC experiments gets modified in the NHSSM 
scenario. In section 4 we conclude.
%
\section{The theoretical setup}
\label{sec:setup}
%
In this section, we first briefly discuss the ingredients of the NHSSM scenario which can be seen as an augmented version of the much popular MSSM scenario. Given that the bottom and the tau Yukawa couplings are central to the present study, we then outline how the radiative corrections to the running masses (Yukawa couplings) $m_{b,\tau}$ ($y_{b,\tau}$) get modified in the NHSSM scenario when compared to the corresponding MSSM contributions to the same.
Finally, we touch upon the structure of the Higgs sector which, by itself, is no different from that in the MSSM scenario. Further, we point out a few of its salient features that are in the direct context of the present study.
\subsection{The NHSSM scenario}
\label{subsec:nhssm}
The Lagrangian of the NHSSM scenario involving the soft SUSY breaking terms includes, in addition to those that are already present in the
MSSM, nonholomorphic terms involving the scalars ($\phi^2 \phi^*$) and the
fermions ($\psi \psi$). These are given by~\cite{Martin:1999hc, Chattopadhyay:2016ivr}
\begin{eqnarray}
\label{nh_lagrangian}
 -\mathcal{L'}_{\text{soft}}^{\phi^{2}\phi^{*}} &=& \tilde{q} \cdot h_{d}^* \,
 A_{u}' \, \tilde{u}^* + 
 \tilde{q} \cdot h_{u}^* \, A_{d}' \, \tilde{d}^* +\tilde{\ell}\cdot h_{u}^* \, A_{\ell}' \, \tilde{e}^* + \mathrm{h.c.} \; , \\
 -\mathcal{L'}_{\text{soft}}^{\psi\psi} &=& {\mathbf \mu^{\prime}} \tilde{h}_u\cdot
\tilde{h}_d \; ,
\end{eqnarray}
where $\tilde{q}$, $\tilde{u}$ and $\tilde{d}$ represent the left-handed 
(doublet) squark states, the right-handed (singlet) up- and down-type squark 
states, respectively, while $\tilde{\ell}$ and $\tilde{e}$ stand for the left-handed (doublet) 
and the right-handed (singlet) slepton states. $h_u$ and $h_d$ are the Higgs fields as introduced in the previous section
while $\tilde{h}_u$ and $\tilde{h}_d$ are the corresponding higgsino states. The trilinear parameters $A_{f=u,d,\ell}'$ are defined 
as $A'_f= y_f \, {\cal A}'_f$\footnote{The choice of the parameter $A_f'$ defined this way proves to be convenient as the numerical packages that we employ later in our analysis adopt the same. This is true for the trilinear parameters, $A_f$, of the MSSM as well, i.e., $A_f=y_f {\cal A}_f$.} with `$f$' denoting the type of 
the squark or the slepton it is related to, $y_f$ and ${\cal A}'_f$ being the Yukawa coupling of the corresponding 
quark/lepton and its associated NH soft trilinear parameter without scaling with the Yukawa coupling, respectively. It should be noted that in contrast to the MSSM case, conjugate Higgs fields $h_d^*$ and $h_u^*$ now appear in $\mathcal{L'}_{\text{soft}}^{\phi^{2}\phi^{*}}$ and to respect the hypercharge assignments these now couple to up-type squarks and down-type squarks/charged sleptons, respectively. $\mu'$ represents the higgsino mass parameter of 
the NHSSM which is a soft SUSY-breaking parameter of the scenario, in contrast 
to its counterpart `$\mu$' in the MSSM which appears in the holomorphic,
(SUSY-conserving) superpotential term $\mu \tilde{h}_u\cdot \tilde{h}_d$.

The presence of the NH soft trilinear terms involving ${\cal A}'_f$ 
modifies the off-diagonal terms of the tree-level sfermion mass-squared 
matrices of the MSSM. Thus, in the NHSSM, these modified elements have 
the generic form
$-y_f \, v \left[{\cal A}_f-(\mu+{\cal A}'_f) r_\beta \right]$
for the up- (with $r_\beta=\cot\beta$) and the down-type
squark/slepton (with $r_\beta=\tan\beta$) sectors \cite{Chattopadhyay:2016ivr} 
with $v=\sqrt{v_u^2+v_d^2} =246$ GeV.
Note that unlike in the MSSM, contributions from the NH soft trilinear
parameters ${\cal A}'_f$ arise through their products with 
either $\cot\beta$ (for the up quark sector) or $\tan\beta$ (for the
down quark and the charged lepton sectors). Thus, for large $\tanb$, 
such contributions might be significant in the bottom and the tau sectors. In 
turn, any phenomenological observable that involves a chiral mixing of 
sfermions would be affected by such NH parameters.

As for the neutralino and the chargino (the electroweakinos) sectors, those are 
affected at the tree level in the presence of the NH $\mu'$-term. 
There, the modifications appear as shifts $\mu \to \mu+\mu'$ in the 
entries of the higgsino blocks of the respective mass
matrices~\cite{Chattopadhyay:2016ivr}.
%
\subsection{Radiative corrections to $y_b$ and $y_\tau$}
\label{subsec:radcor-yb-ytau}
%
An important implication of the presence of NH soft trilinear terms involving 
$A_f^\prime$ ($\abprime$ and $\atauprime$) in the NHSSM Lagrangian is that 
those could influence the radiative corrections to the down-type Yukawa 
couplings, $y_b$ and $y_\tau$, thus modifying their tree-level relations to 
$m_b$ and $m_\tau$, respectively. These are over and above their dependencies 
on $\tanb$ that are already present in the MSSM.

At one-loop level, the principal MSSM contributions to $y_b$ arise from the 
gluino-sbottom ($\tilde{g}-\tilde{b}_i$) loop and the chargino-stop
($\charpm_i-\tilde{t}$) loop
\cite{Hall:1993gn, Hempfling:1993kv, Carena:1994bv, Pierce:1996zz,
Logan:2000cz, Antusch:2008tf}.%
\footnote{The SUSY-QCD and the SUSY-electroweak
corrections can have enhanced contributions for moderate to large `$\mu$' and ${\cal A}_t$ when $\tanb$ is large. An improved perturbative result is found by resumming these
dominant contributions which is also extended to include the effect from large ${\cal A}_b$
that is absent at one-loop level
\cite{Carena:1999py, Guasch:2003cv}. Given that the parameters $\alphas$ and $y_t$ that enter the corrections
have significant dependence on the renormalization scale,
a further improvement in the estimation with much reduced uncertainty is found by the inclusion of two-loop SUSY-QCD corrections
\cite{Noth:2008tw, Noth:2010jy, Mihaila:2010mp, Ghezzi:2017enb}.
}
On the other hand, in the NHSSM where $H_u$ is 
also associated with the down-type squarks (and sleptons) in the trilinear soft 
interactions, one can have an extra $\tilde g-\tilde b$ contribution with a 
factor $y_b \, {\cal A}'_b$ showing up at the $\tilde{b}_L$-$H_u^*$-$\tilde{b}_R$ vertex. 
Besides, there is a neutralino-sbottom ($\ntrli-\tilde b$) loop contribution 
with the same factor $y_b \, {\cal A}'_b$ appearing at the trilinear scalar vertex and a factor 
of $y_b^2$ coming from the other two vertices having $\sbleft \sbright \chi^0$ 
interactions with their origins in the superpotential \cite{Chattopadhyay:2018tqv}. 
At one-loop level, the modified relation between $m_b$ and $y_b$ in the NHSSM 
(which is identical in appearance to that in the MSSM)\footnote{References 
\cite{Crivellin:2010er, Crivellin:2011jt, Crivellin:2012zz} contain detailed analysis 
on the effective Higgs boson vertices that involve nonholomorphic ${\cal A}'_f$ terms in 
SUSY-QCD and electroweak corrections to $m_f$
($y_f$).} and hence the radiatively-corrected expression for $\yb$ are given by
\beq
m_b = \frac{y_b v_d}{\sqrt 2} (1+\Delta_b) \quad \Longrightarrow
\quad
\yb = \frac{\sqrt{2} m_b}{v_d (1+\Delta_b)} \, ,
\label{eqn:yb-correction}
\eeq
where $\Delta_b$ is the SUSY threshold correction which now receives, in 
addition to the MSSM contributions, the NHSSM ones and are to be evaluated at 
the scale $M_{\text{SUSY}}$ given by the typical mass of the SUSY excitations 
that run in the loops. For $\tanb >>1$ and $M_\text{SUSY} >>m_Z$, $\Delta_b$ 
is given by\footnote{In the presence of an $\alpha_s$ driven contribution and contributions that are enhanced for large values of
$|{\cal A}'_b|$ and $|{\cal A}_t|$ (that we require to have the mass of the
SM-like Higgs boson in the right ballpark), we have dropped from the expression of $\Delta_b$ the contributions
proportional to the $U(1)$ and $SU(2)$ gauge couplings, $g'$ and $g$. However,
all possible contributions to such corrections are taken care of by the
numerical package we use for the purpose (see section
\ref{subsec:bb-tautau-phi}).}
\bea
\Delta_b \, &\simeq \,
\frac{y_{t_{\mathrm{SM}}}^2}{16 \pi^2} \, \mu'' {\cal A}_t \, I( \mstone^2, \msttwo^2, \mu''^2) \tan\beta +
\frac{2\alphas}{3\pi} \, \mgluino \, (\mu + {\cal A}'_b)
 \, I( \msbone^2, \msbtwo^2, m_{\tilde g}^2)\tan\beta
\nonumber \\
 &\hskip 152pt + \frac{y^2_{b_{\mathrm{SM}}}}{16 \pi^2} \, \mu'' \, (\mu+{\cal A}'_b) \, I( \msbone^2, \msbtwo^2, \mu''^2) \tan\beta \, ,
\label{eqn:deltab}
\eea
where $\mu''=\mu+\muprime$ and the characteristic loop (integral) function $I(a,b,c)$ involving loop-momentum `$u$' and squared masses $a, b, c$ of the three propagating states in the loop is given by \cite{Hall:1993gn}
\begin{equation}
I(a,b,c) = \int_0^\infty \frac{udu}{(u+a)(u+b)(u+c)}
         = \frac{ab~\ln(a/b)+bc~\ln(b/c)+ca~\ln(c/a)}{(a-b)(b-c)(a-c)} \, ,
\label{eqn:I-value}
\end{equation}
and $\alphas$ is the strong coupling constant and ${\cal A}_t$ and
$\mgluino$ are the soft SUSY breaking holomorphic (of the MSSM) trilinear coupling of the top squarks to 
the up-type Higgs field (without a scaling with $y_t$) and the mass of the gluino, respectively.
$m_{\tilde{t}_{1(2)}}$ and $m_{\tilde{b}_{1(2)}}$ stand for the masses of the lighter (heavier) top and bottom squark states, respectively. To make a generic numerical sense of this involved function, it may be noted that the same can be parametrized as $I(a,b,c)=K_I/a_{max}$ \cite{Carena:1994bv} where
$a_{max} = \mathrm{max}(a,b,c)$ is the maximum of the three squared masses appearing there and $K_I \simeq 0.5-0.9$ in the absence of a large hierarchy between these  masses.\footnote{The limiting values of $K_I =0.5 \, (1)$ are obtained by carrying out the shown integration with one (two) of the masses vanishing.}

The corresponding corrections to $y_\tau$ ($m_\tau$) come from $\charpm_i$--$\tilde{\nu}_\tau$ and $\ntrli$--$\tilde{\tau}$ loops. Similar to the case of $\Delta_b$, $\Delta_\tau$ is now driven by ${\cal A}'_\tau$ and $\tanb$ where, unlike in the former case, roles played by $g'$ and `$g$' are important, in general. Thus, the relations in equations \ref{eqn:yb-correction} and \ref{eqn:deltab} can be used for the modified expression for $m_\tau$ ($y_\tau$) by dropping first the
$\alpha_{_S}$-dependent gluino contribution straightaway and replacing all bottom(top)-flavor related parameters by  their leptonic ($\tau$($\nu_\tau$)) counterparts before including the $g'$- and $g$-dependent terms. As a result, the corresponding first term which would represent the
contribution from the $\charpm_i$--$\tilde{\nu}_\tau$ loop and would go as the (vanishing) neutrino Yukawa coupling ($y_t \rightarrow y_{\nu_\tau}$) also drops out. Tweaking appropriately the expressions for the $g'$- and
$g$-dependent MSSM contributions as presented in reference \cite{Girrbach:2009uy} for the NHSSM scenario,
$\Delta_\tau$ takes the following form:
%
\begin{align}
\Delta_\tau &\simeq
\frac{y^2_{\tau_{\mathrm{SM}}}}{16 \pi^2} \, \mu'' \, (\mu+{\cal A}'_\tau) \, I(\mstaul^2, \mstaur^2, \mu''^2) \tanb
+ \frac{g'^2}{16 \pi^2} \, \mu'' \, (\mu+{\cal A}'_\tau) \, I(\mstaul^2,\mstaur^2, \mu''^2) \tanb \nonumber \\
& + \frac{g'^2}{16 \pi^2} \, \mone \, 
\left[ \mu'' \left\{ {1 \over 2} I(\mone^2, \mu''^2, \mstaul^2)
- I(\mone^2, \mu''^2, \mstaur^2) \right\} + \mu \, I(\mone^2, \mstaul^2, \mstaur^2) \right] \tanb \nonumber \\
&- \frac{g^2}{16 \pi^2} \, \mtwo  \, \mu'' \, \left[ {1 \over 2} I(\mtwo^2, \mu''^2, \mstaul^2)  + I(\mtwo^2, \mu''^2, m_{\tilde{\nu}_{\tau}}^2) \right] \tanb \, .
\label{eqn:deltatau}
\end{align}
%

\noindent
The modified relation between $m_\tau$ and $\ytau$ and hence the expression for corrected $\ytau$ are given by 
\beq
m_\tau = \frac{\ytau v_d}{\sqrt 2} (1+\Delta_\tau) \quad \Longrightarrow
\quad
\ytau = \frac{\sqrt{2} m_\tau}{v_d (1+\Delta_\tau)} \, .
\label{eqn:ytau-correction}
\eeq

As can be seen from equations \ref{eqn:yb-correction} -- \ref{eqn:ytau-correction}, 
the one-loop corrected $y_{b(\tau)}$ not only has an explicit direct dependence on $\tanb$ as is the case in the MSSM but also has an additional  dependence on ${\cal A}'_b \, ({\cal A}'_\tau)$ (via the quantity $\mu+{\cal A}'_b$ \, ($\mu+{\cal A}'_\tau$))
in the NHSSM which is roughly on the same 
footing as $\tanb$ when `$\mu$' is not too large.
Furthermore, it is clear from the expressions in equations \ref{eqn:yb-correction} and \ref{eqn:ytau-correction} that one expects enhancements (suppressions) in $y_{b,\tau}$ (over their usual variations with $\tanb$) for $\Delta_{b,\tau} < 0 \, (> 0)$.
%
\subsection{The Higgs sector}
\label{subsec:higgs-sector}
%
As mentioned earlier, the Higgs sector of the NHSSM is structurally similar to that of the MSSM having two complex Higgs doublets of opposite hypercharges thus
leading to five physical Higgs states once the electroweak symmetry gets broken.
These are the two $CP$-even Higgs states, `$h$' and `$H$', of which the lighter one ($h$) is SM-like ($h \equiv \hsm$); a $CP$-odd one, `$A$', and two charged
Higgs states, $H^\pm$.

At the tree level, the masses of these physical states and their couplings with the SM fermions and the gauge bosons depend only on two free parameters. 
These are conventionally chosen to be $\tanb$  and the mass of the physical $CP$-odd neutral Higgs state, $m_A$ \cite{Djouadi:2005gj}. In terms of these two parameters, the masses of the other four Higgs states, at the tree level, are given by
\bea
m_{h,H}^2 &=& {1 \over 2} \bigg[ m_A^2+m_Z^2 \mp \sqrt{(m_A^2+m_Z^2) - 4 m_A^2 m_Z^2 \cos^2 2\beta} \bigg], \quad m_{H^\pm}^2 = m_A^2 + m_{W^\pm}^2 \; ,
\eea
where $m_Z$ and $m_{W^\pm}$ are the masses of the SM `$Z$' and $W^\pm$ bosons, respectively.
It follows that for $m_A >> m_Z$, $m_H \approx m_{H^\pm} \approx m_A$.
When the Higgs sector conserves the $CP$ symmetry, the physical $CP$-even neutral Higgs states ($h,H$) are obtained from their weak counterparts
via the rotation~\cite{Gunion:1984yn, Gunion:1989we}
\beq
\label{eq:hH2HDM}
\begin{pmatrix}
H \\ h
\end{pmatrix}
=
\begin{pmatrix}
 \cos\alpha \; & \sin\alpha \\
-\sin\alpha \; & \cos\alpha
\end{pmatrix}
\begin{pmatrix}
H_1^0 \\ H_2^0
\end{pmatrix} \, ,
\eeq
where the Higgs mixing angle `$\alpha$' controls the interactions of these physical Higgs states with the 
fermions, the gauge bosons, the sfermions and the  electroweakinos and is given by $m_A$ and $\tanb$ as
\beq
    \alpha
  = \frac{1}{2} \tan^{-1} \left( \tan 2 \beta \, \frac{m_A^2 + m_Z^2}{ m_A^2 - m_Z^2} \right) \quad \mathrm{with} \quad
   - \frac{\pi}{2} \leq \alpha \leq 0 \ .
\label{alpha:tree}
\eeq
The couplings of these Higgs states with the bottom quark, which are in direct reference in the present work, are given at the tree level by their MSSM expressions \cite{Djouadi:2005gj}
\begin{align}
G_{hbb} &= -i {m_b \over v} {\sin\alpha \over \cos\beta} \; , \qquad
G_{Hbb} = i {m_b \over v} {\cos\alpha \over \cos\beta} \; , \qquad 
G_{Abb} = {m_b \over v} \tanb \, \gamma_5  \; .
\label{eqn:hbb-couplings}
\end{align}
In the units of the tree-level SM $b$-quark Yukawa coupling $y_b={i m_b \over v}$, the first two couplings shown in equation \ref{eqn:hbb-couplings} reduce to
\bea
g_{hbb} &=& -{\sin\alpha \over \cos\beta} = \sin(\beta-\alpha) - \tanb \cos(\beta-\alpha) \nonumber \; ,\\
g_{Hbb} &=& {\cos\alpha \over \cos\beta} = \cos(\beta-\alpha) + \tanb \sin(\beta-\alpha) \, .
\label{eqn:hbb-reduced}
\eea
Thus, in the so-called decoupling limit ($m_A >> m_Z$ and $\tanb >>1$), when
$\cos(\beta-\alpha) \to 0$, $\sin(\beta-\alpha) \to 1$ and the factor $\gamma_5$ is ignored,\footnote{For a massive $b$-quark, the process of dimensionally regularizing the calculation is somewhat involved given that the Dirac matrix
$\gamma_5$ is a four-dimensional construct. The resulting effects of a finite $m_b$ are small $\sim {\cal O}\left({m_b^2 \over m_A^2}\right)$ \cite{Dawson:2005vi}.} one finds
\beq
g_{hbb} \to 1, \qquad  g_{Hbb} \to \tanb, \qquad g_{Abb} = \tanb \; .
\eeq
The bottom line is that the $Hbb$ coupling in the decoupling limit and the $Abb$ coupling
are both enhanced by a factor of $\tanb$. On the other hand, the $hbb$ coupling gets $\tanb$-enhanced in the non-decoupling limit for 
which $m_A \gtrsim m_Z$. It suffices to mention here that, in contrast, the 
coupling of the Higgs bosons to the top quark(s) are generically
$\tanb$-suppressed (i.e., $\cot\beta$-enhanced) and the couplings of `$h$' and `$H$' to 
top quarks have just the opposite dependencies on $\tanb$ when compared to the 
same to bottom quarks.

At this point, it is important to note that these couplings receive crucial higher-order (SUSY-QCD and SUSY-electroweak) corrections whereby the reduced Yukawa 
couplings factors with resummed contributions (as discussed in section \ref{subsec:radcor-yb-ytau}) included, for the neutral Higgs states, are given by
\cite{Carena:2002es, Carena:1999py, Guasch:2003cv}
\bea
g_{hbb} &\simeq& -{\sin\bar{\alpha} \over \cos\beta} \, \Big[ 1- {\Delta_b \over {1+\Delta_b}} \, (1+\cot\bar{\alpha} \cot\beta) \Big] \nonumber \; ,\\
g_{Hbb} &\simeq& {\cos\bar{\alpha} \over \cos\beta} \, \Big[ 1- {\Delta_b \over {1+\Delta_b}} \, (1-\tan\bar{\alpha} \cot\beta) \Big] \nonumber \; ,\\
g_{Abb} &\simeq& \tanb \, \Big[ 1- {\Delta_b \over {1+\Delta_b}} \, {1 \over \sin^2\beta} \Big] \; ,
\label{eqn:couplings}
\eea
where $\bar{\alpha}$ stands for the radiatively corrected Higgs mixing angle which now includes possible NHSSM effects
in addition to the ones of the MSSM origin and $\Delta_b$ is as defined in equation \ref{eqn:deltab}.

An analogous $\tanb$-dependence follows for $y_\tau$ but for the absence of an
$\alpha_s$-driven gluino contribution to $\Delta_\tau$ (as given in equation
\ref{eqn:deltatau}; when compared to equation $\Delta_b$ of \ref{eqn:deltab}). Thus, $\ytau$ would be intrinsically smaller and so would be its range of variation about its nominal MSSM value when $\atauprime$ is varied, as opposed to
the range of variation of $\yb$ under a varying $\abprime$. This implies that it is
$\abprime$, rather than $\atauprime$, that would broadly dictate the decay branching fractions of $H/A$ to $\tautaubar$, the final state which the LHC experiments exploit in their search for these Higgs states in their associated productions with a $\bbbar$ pair \cite{Aad:2020zxo,Aaboud:2017sjh,Sirunyan:2018zut}.

Armed with the above knowledge of the $(H/A) \, \bbbar$, $(H/A) \tautaubar$ couplings in the NHSSM, we now take up the study of the production of these Higgs states in association with a $\bbbar$ pair at the LHC and their subsequent decays to $\bbbar$ and $\tautaubar$ pairs. The goal is to find out how the variation of the quantity
$\sigma_{(pp \to \bbbar H/A)} \times \mathrm{BR}[H/A \to \tautaubar]$ that offers
the most stringent LHC exclusion of the customary $m_A-\tanb$ plane differs in the NHSSM
from that in the MSSM in the presence of a non-vanishing $\abprime$ and hence the impact of the same in modifying the said exclusion region.
In the subsequent sections of this work where we present and discuss our results, unless otherwise mentioned, we would refer to the trilinear parameters $\abprime$ and $\atauprime$ as defined in section~\ref{sec:setup}.
%
\section{Results}
\label{sec:results}
%
In this section, we present our results by first studying the dependencies of the 
Yukawa couplings $y_b$ and $y_\tau$ on the NHSSM trilinear parameters
$\abprime$ and $\atauprime$, respectively, and $\tanb$. While for the 
productions of `$H$' and `$A$' in association with a $\bbbar$ pair, variations of both $y_b$ and $\tanb$ play major roles, their decays, 
are expected to be further governed by $y_\tau$. A study on the decays of 
the heavy Higgs bosons of the NHSSM follows. We then touch upon some important 
theoretical issues in the calculation of the cross section for the process
$pp \to \bbbar H/A$ at the LHC energies before describing our approach of calculating 
the same. All these are aimed to find out how the nontrivial 
dependencies of 
various (Yukawa) couplings of such Higgs bosons affect the sensitivity of the LHC to 
these states when produced in association with a pair of bottom quarks. This then 
prompts us to examine to what extent the relevant constraints obtained from the latest LHC searches in 
the said mode, for an MSSM scenario, could get relaxed or further strengthened in the NHSSM scenario of our interest.
For the numerical analysis, we employ the publicly available implementation of the NHSSM model\footnote{Some useful patches are kindly provided by the authors of {\tt SARAH}.} in {\tt SARAH-v4.14.4} \cite{Staub:2013tta, Staub:2015kfa}-generated {\tt SPheno} \cite{Porod:2011nf}.\footnote{Note that {\tt SARAH}-generated {\tt SPheno} follows the convention
$A_{f,f'} \equiv y_f {\cal A}_{f,f'}$ which is adopted by us and is discussed earlier in section \ref{subsec:nhssm}.}
Unless otherwise mentioned, all subsequent figures make use of the fixed values and the ranges of variation (provided at the scale $M_{\text{SUSY}} \simeq 3$ TeV) of various SUSY input parameters of Table \ref{tab:inputs}. These conform to  $122\, \mathrm{GeV}<\mhsm < 128\, \mathrm{GeV}$ and pass the constraints from {\tt HiggsBounds-v5.9.1} \cite{Bechtle:2020pkv}.

To focus on the NHSSM-specific sector of the parameter space, in particular, 
the part that involves the trilinear coupling parameters $\abprime$ and $\atauprime$ 
and which could affect the phenomenology of the bottom quark and tau lepton, we fix all other SUSY parameters to their suitable values such that various important bounds 
from the LHC experiments are complied with and the observed properties of the SM-like 
Higgs bosons could be ensured. We treat $\tanb$ as a free parameter since 
its interplay with $\abprime$ and $\atauprime$ is central to our present study. 

Note that we fix `$\mu$' to not too large a value of 500 GeV so that the scenario remains somewhat `natural'. Given that the contributions to $\Delta_{b,\tau}$ depend significantly on the quantity $\mu+{\cal A}'_{b,\tau}$, we would thus look out for relatively large $|\abtauprime|$ to find $\Delta_{b,\tau}$ to be substantially large. Furthermore, we choose a somewhat large value of $\muprime$ (1.5 TeV) such that the higgsino-like electroweakinos have masses around $\sim \mu+\muprime = 2$ TeV. Together with $\mone$ and $\mtwo$ set to 2 TeV, these would ensure that the electroweakinos are heavy enough not only to pass the current experimental bounds on them but also to render themselves inaccessible for the `$H$' and the `$A$' states decaying to a pair of them.
%
\begin{table}[t]
\centering
\begin{tabular}{|c|c|c|}
\hline\hline 
Parameters & MSSM & NHSSM \\ [0.5ex]
\hline
$\mu$ (GeV)
& \multicolumn{2}{|c|}{500} \\
$\mone, \mtwo, \mthree$ (TeV)
& \multicolumn{2}{|c|}{2, 2, 3} \\
$\tan\beta$
& \multicolumn{2}{|c|}{[5 : 60]} \\
$m_A$ (TeV)
& \multicolumn{2}{|c|}{[0.35 : 3]} \\
$m_{\tilde{q}}, m_{\tilde{\ell}}$ (TeV)
& \multicolumn{2}{|c|}{3} \\
$A_{t,b,\tau}$ (TeV) & \multicolumn{2}{|c|}{-2.7, 0, 0} \\
\hline
$\muprime$ (TeV) & -- & 1.5 \\
$A_t^\prime$ (TeV) & -- & 0  \\
$\abprime, \atauprime$ (TeV) & -- & [-3 : 3]  \\
\hline
\end{tabular}
\caption{Fixed values and ranges adopted for various input SUSY parameters provided to {\tt SPheno} at the scale $M_{\text{SUSY}} \simeq 3$ TeV.
}
\label{tab:inputs}
\end{table}
%
\subsection{Variations of $bb\Phi$ and $\tau\tau\Phi$ interaction strengths}
\label{subsec:bb-tautau-phi}
%
The characteristic dependencies of $y_b$ and $y_\tau$ and the same for the 
resulting strengths of the $\Phi bb$ and $\Phi \tau\tau$  interactions, respectively, where 
$\Phi \in \{H,A\}$, are discussed in detail in the previous section. In 
particular, note that the corrections to $y_{b,\tau}$ and the very 
form of the interactions of these heavy Higgs bosons to a bottom quark or a tau 
lepton pair in the decoupling regime each lends a separate $\tanb$-dependence to 
these couplings (see equation \ref{eqn:couplings}). In addition, as discussed in section \ref{sec:setup}, in the NHSSM, 
corrections to $m_{b,\tau}$ ($y_{b,\tau}$) depend on $\abtauprime$. In this 
subsection we study these dependencies on $\tanb$ and $\abtauprime \, (=y_{b,\tau} \,  {\cal A}'_{b,\tau})$.

%

%
\begin{figure}[t]
\centering
\subfigure[]{
\includegraphics[width=0.425\textwidth,height=0.245\textheight]{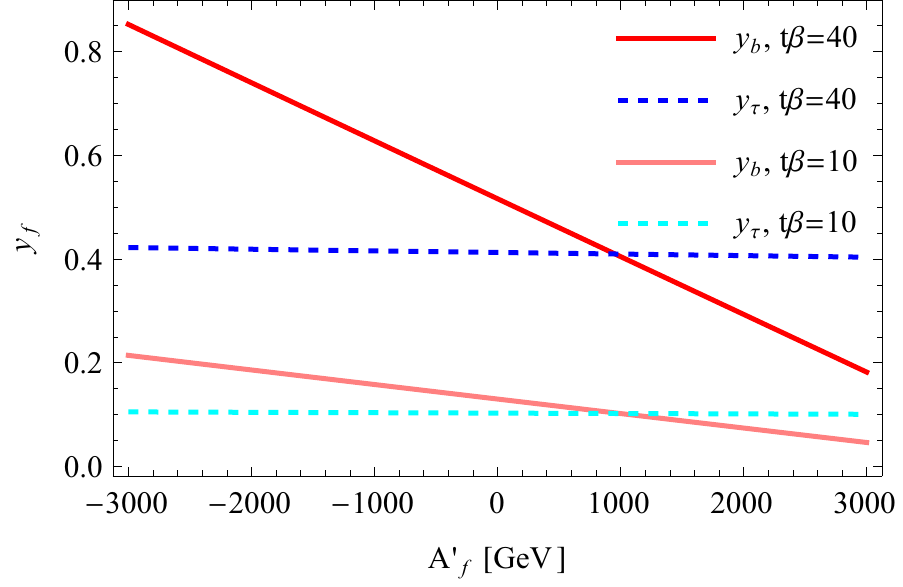}
\label{subfig:yf-afp-tb}
}
\hskip 25pt
\subfigure[]{
\includegraphics[width=0.425\textwidth,height=0.245\textheight]{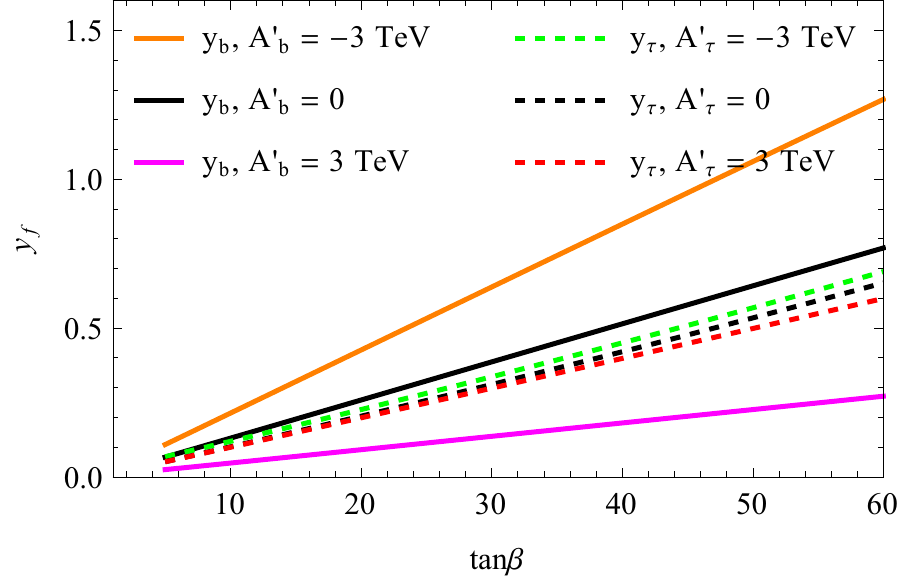}
\label{subfig:yf-tb-afp}
} \\
\vskip 20pt
\hskip 30pt
\subfigure[]{
\includegraphics[width=0.40\textwidth, height=0.26\textheight]{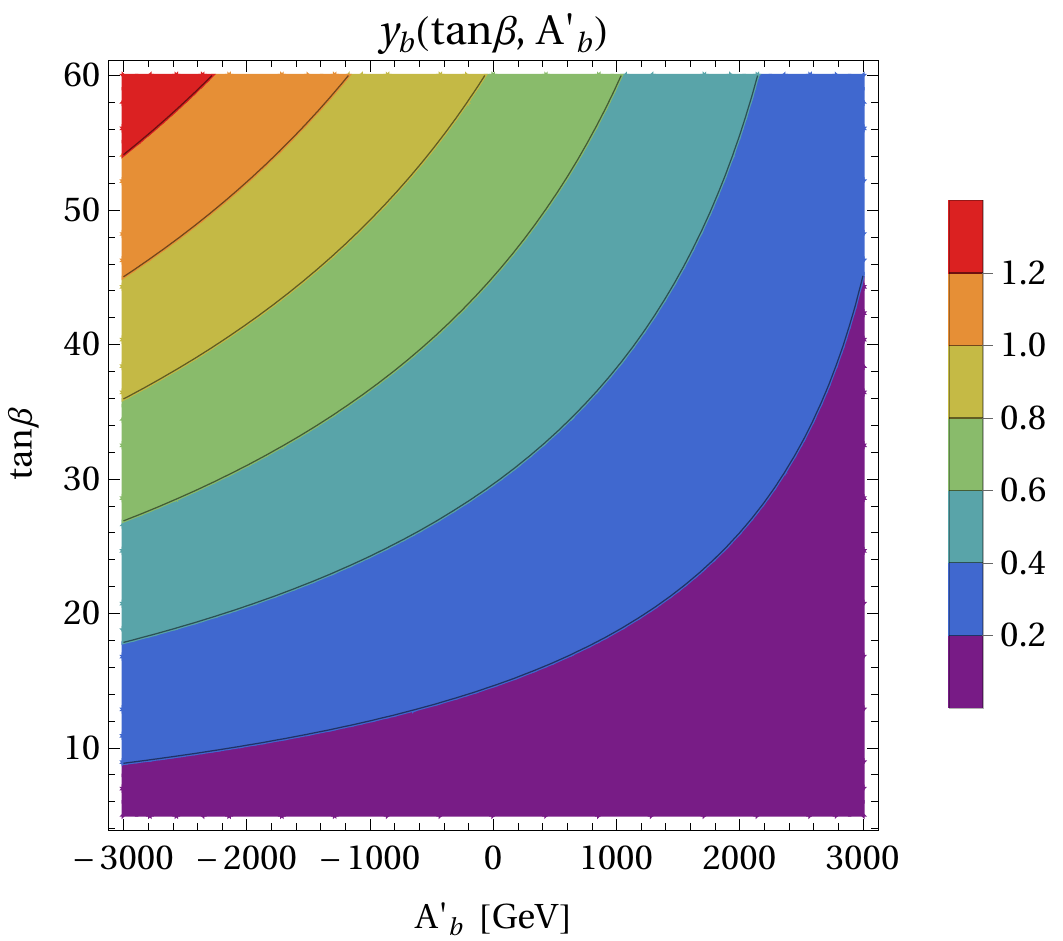}
\label{subfig:yb-abp-tb}
}
\hskip 40pt
\subfigure[]{
\includegraphics[width=0.40\textwidth, height=0.26\textheight]{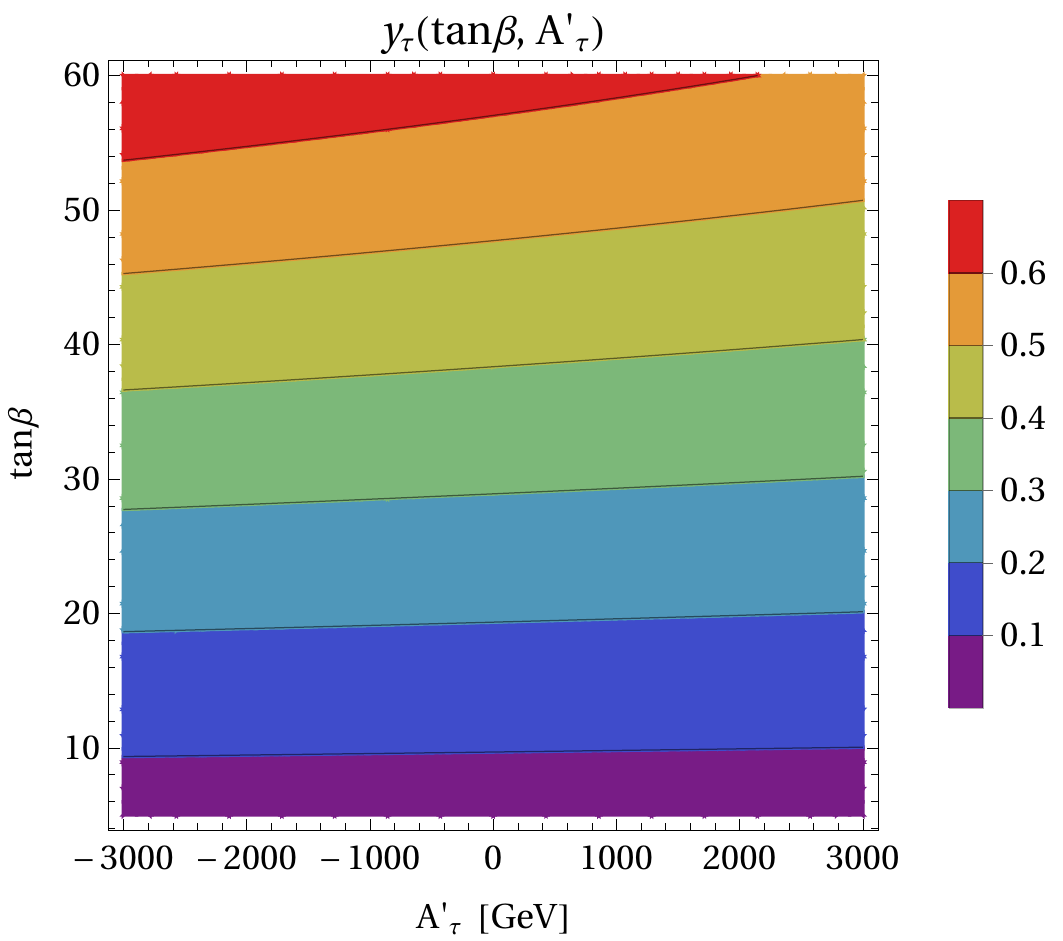}
\label{subfig:ytau-atau-tb}
}
\caption{Variations of $y_{f=b,\tau}$ as functions of (a) $A'_{f=b,\tau}$, for two fixed values of $\tanb$ (10 and 40) and of (b) $\tanb$, for three fixed values of $A'_{f=b,\tau}$  ($-3$ TeV, 0 and 3 TeV). Variation of $\yb$ ($\ytau$) in the $\abprime (\atauprime)-\tanb$ plane is shown in plot (c) ((d)).
}
\label{fig:yb-tanb-abprime}
\end{figure}
First, we discuss the dependence of $\ybtau$ ($\mbtau$) on $\abtauprime$ and
$\tanb$ arising from radiative corrections to $\mbtau$. 
As has been noted at the end of section \ref{subsec:radcor-yb-ytau}, a positive 
(negative) $\Delta_{b,\tau}$ turns the radiatively corrected values of
$y_{b,\tau}$ smaller (larger) than their nominal values. In both 
cases, effects are amplified for larger values of $\tanb$. In figure 
\ref{subfig:yf-afp-tb} we illustrate the variations of $y_{b,\tau}$ as 
functions of $A'_{b,\tau}$ for two different values of $\tanb$ (=10 and 
40). The corresponding MSSM values of $y_{b,\tau}$, for any given value 
of $\tanb$, are closely the ones found at the point of intersection of the particular 
curve (line) and the line erected vertically at $A'_{b,\tau}=0$.
On the other hand, figure \ref{subfig:yf-tb-afp} shows variations $\yb$ and
$\ytau$ as functions of $\tanb$ for three different values $A'_{b,\tau}$
($=-3, 0, 3$ TeV). Again, lines with $A'_{b,\tau}=0$ (in black) 
would closely correspond to the variations of $y_{b,\tau}$ with $\tanb$ in 
the MSSM. These two plots reveal the individual magnitudes of $\yb$ and $\ytau$ 
as they vary and the ranges of their variations. On both counts, $\yb$ wins 
over $\ytau$, hands down, except when $\abprime$ has a large positive value
(3 TeV, in the present case). In the latter case, $\yb$ always remains smaller 
than $\ytau$ for all values of $\tanb$ (see the magenta curve) and their 
difference grows with growing $\tanb$. A much stronger dependence of
$\yb$ on $\tanb$ and $\abprime$, as compared to that of $\ytau$ on $\tanb$ and
$\atauprime$ can be straightaway traced back to the corresponding dependencies 
of $\Delta_b$ and $\Delta_\tau$, respectively, on the relevant pairs from these parameters. These two plots further reveal that while $\yb$ is sensitive to both $\abprime$ 
and $\tanb$ (the sensitivity to the former being milder than that to the 
latter) and assumes larger (smaller) values for large negative (positive)
$\abprime$ and large $\tanb$, $\ytau$ has an almost exclusive and direct 
dependence on $\tanb$ only. Figure \ref{subfig:yb-abp-tb} (\ref{subfig:ytau-atau-tb}) captures the same 
variations but in the plane of $\abprime (\atauprime)-\tanb$.

\begin{figure}[t]
\centering
\subfigure[]{
\includegraphics[width=0.45\textwidth, height=0.25\textheight]{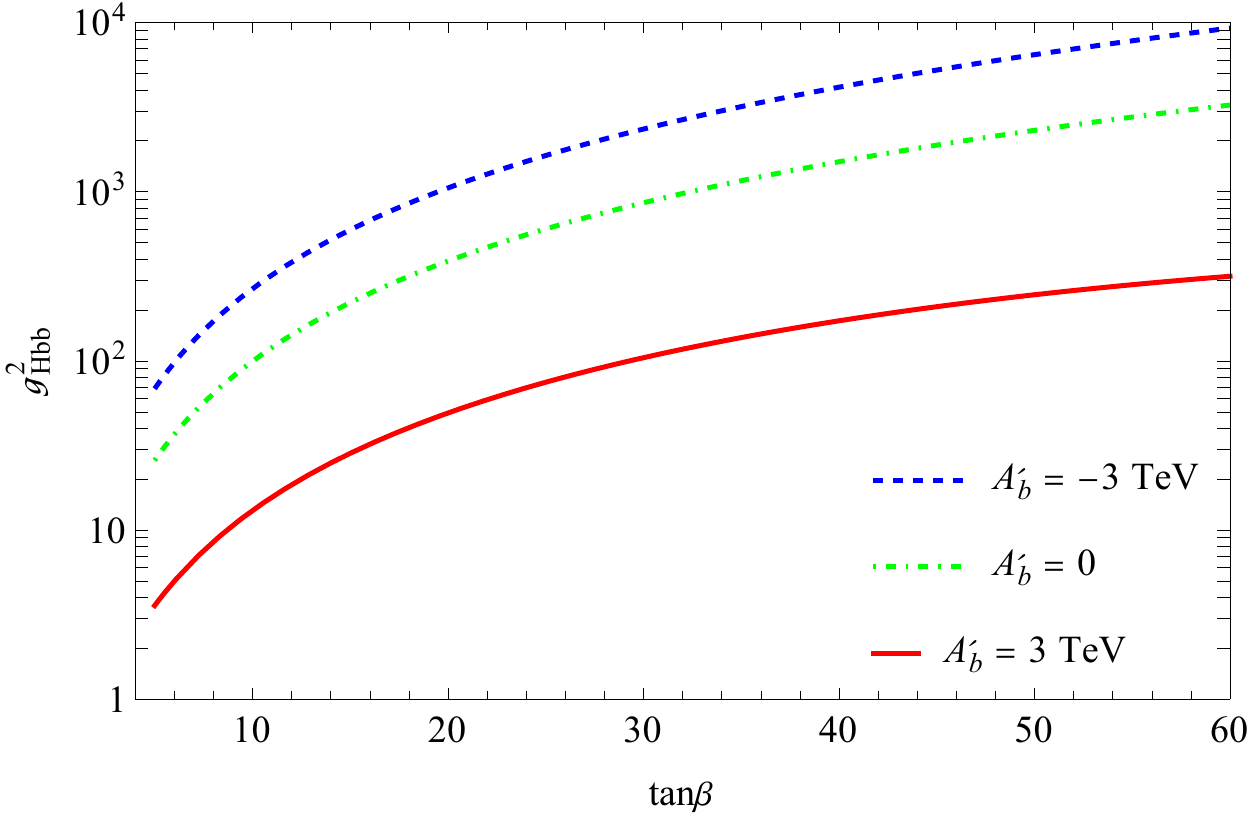}
\label{subfig:gsq-bbphi-tb}
}
\hskip 25pt
\subfigure[]{
\includegraphics[width=0.45\textwidth, height=0.25\textheight]{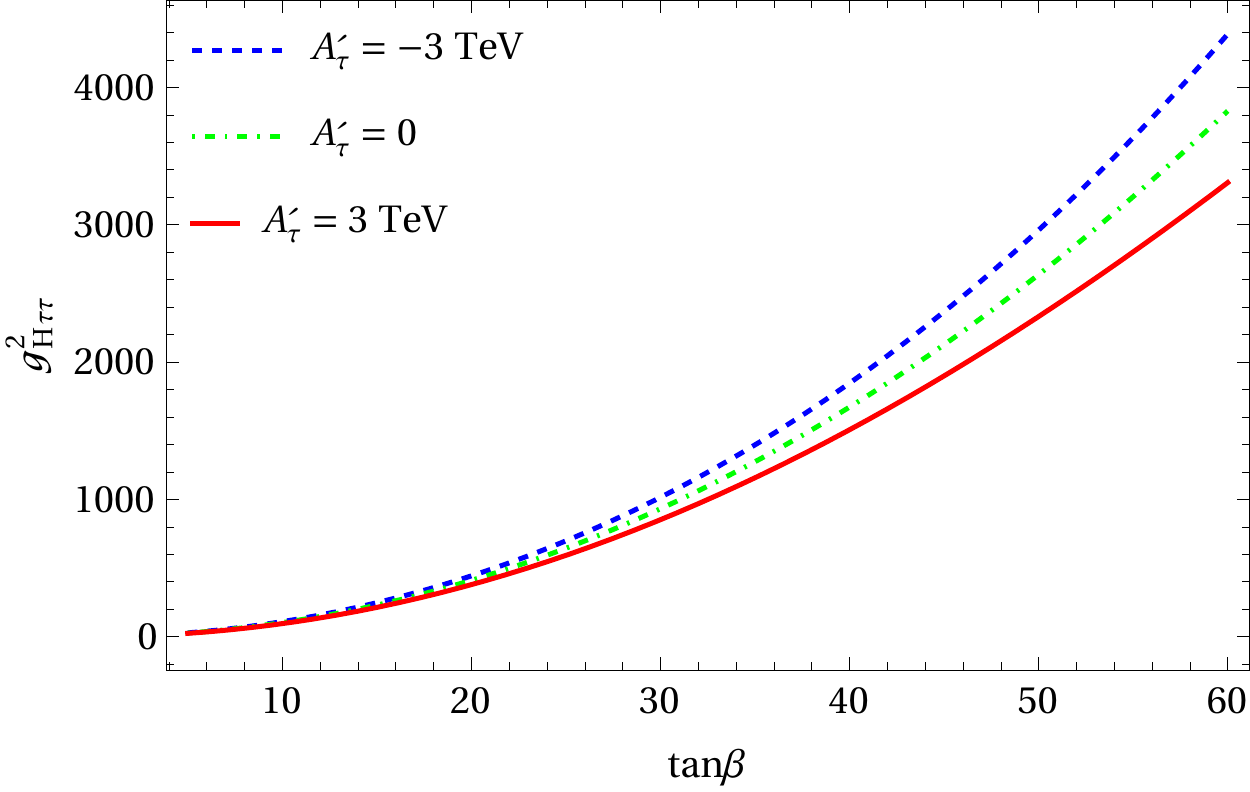}
\label{subfig:gsq-tautauphi-tb}
}
\caption{
Variations of the squared scaling factors $g^2_{Hbb}$ (left) and
$g^2_{H\tau\tau}$ (right) as functions of $\tanb$ and for three fixed values ($-3$ TeV, 0 and 3 TeV) of $\abprime$ (left) and $\atauprime$ (right).
}
\label{fig:gffsqphi}
\end{figure}

The resulting variation of the squared value of the reduced coupling strength for $H bb \, (\tau\tau)$ interaction, i.e., $g^2_{H bb \, (\tau\tau)}$ (see expressions under equation~\ref{eqn:couplings}), in the decoupling regime, as a function of $\tanb$, is shown in
figure~\ref{subfig:gsq-bbphi-tb} (\ref{subfig:gsq-tautauphi-tb}) for $\abprime (\atauprime)=(-3,0,3)$ TeV.
These quantities are extracted from {\tt SPheno} outputs using
equation~\ref{eqn:couplings}.
Note that while we are compelled to adopt a logarithmic vertical axis for figure \ref{subfig:gsq-bbphi-tb} since $g^2_{Hbb}$ takes off rather sharply with a growing $\tanb$ (thanks to a large negative $\abprime$ and a catalyzing role played by the $\alphas$-dependent contribution to $\Delta_b$), we stick to a
linear one in figure \ref{subfig:gsq-tautauphi-tb} to illustrate the corresponding variations in $g^2_{H \tau\tau}$ which are much less sensitive to the signs and magnitudes of $\atauprime$. Curves with $\abprime, \atauprime=0$ roughly illustrate such variations in the MSSM-like setups and are shown for reference purposes.
As has been noted in section \ref{subsec:higgs-sector}, for all 
practical purposes for this work, one can consider
$g^2_{A bb \, (\tau\tau)} =g^2_{H bb \, (\tau\tau)}$. Hence we do not present a 
separate set of plots for $g^2_{A bb \, (\tau\tau)}$.

Given the factor
$g_{H bb \, (\tau\tau)}$ scales the $H bb \, (\tau\tau)$ coupling in the SM to its 
value in the NHSSM, $g^2_{H bb \, (\tau\tau)}$ gives us a quick estimate of the 
factors by which the SM-like values of the associated production cross section
$\sigma_{(pp \to \bbbar H/A)}$ and the relevant decay widths
$\Gamma(H/A \to \bbbar, \tautaubar)$ are to be multiplied to find their corresponding 
values in the NHSSM. In this work we applied such a scaling on the readily available 
values of ${\sigma_{(pp \to \bbbar H/A)}}|_{\mathrm{SM}}$ as a function of $m_{H/A}$ 
(see section \ref{subsec:asso-prod}) to find their corresponding values in the NHSSM. 
On the other hand, although such scalings for $\Gamma(H/A \to \bbbar, \tautaubar)$ 
get automatically taken care of in the {\tt SARAH-SPheno} framework that we use, a 
quantitative knowledge of the magnitudes of $g^2_{H/A \, bb (\tau\tau)}$ and their 
nature of variation over the $\abprime(\atauprime)-\tanb$ plane aid a better 
understanding of the physics interplay in the study of the all important quantity, 
i.e., $\sigma_{(pp \to \bbbar H/A)} \times \mathrm{BR}[H/A \to \tautaubar]$.
%
%
\begin{figure}[t]
\centering
\subfigure[]{
\includegraphics[width=0.45\textwidth, height=0.25\textheight]{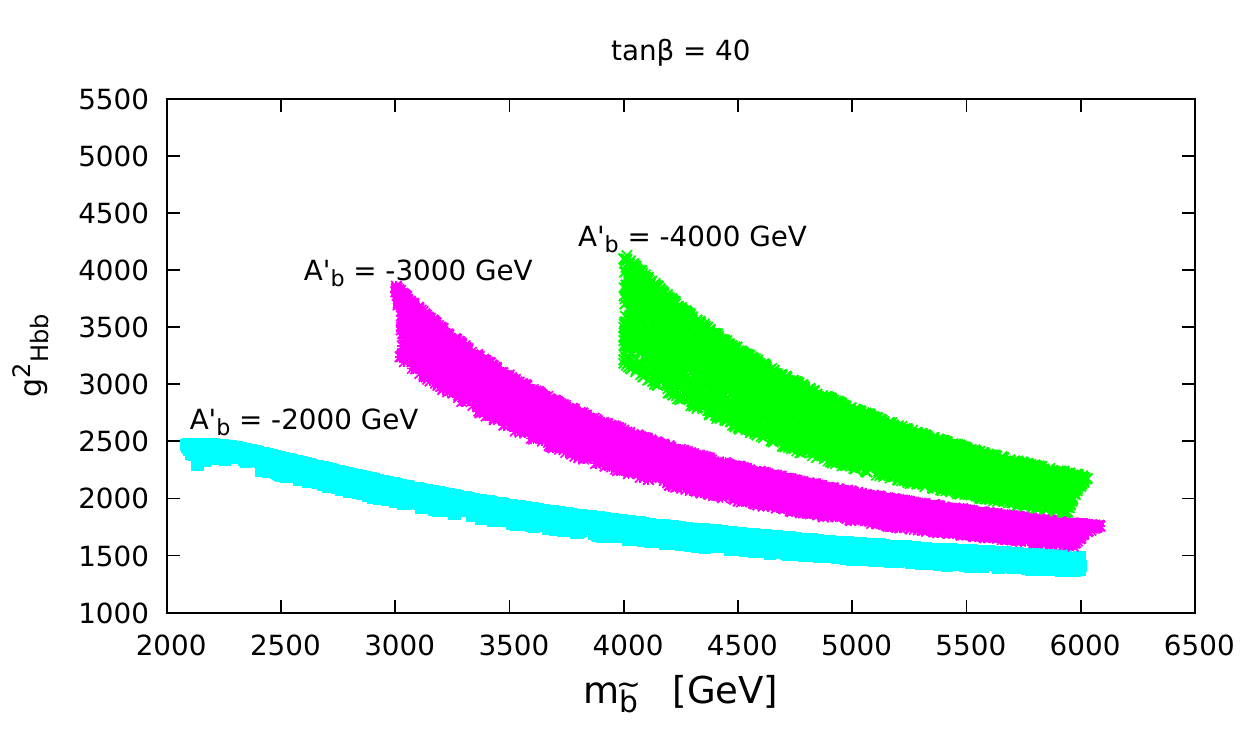}
\label{subfig:msbot-gsq-nabp}
}
\hskip 25pt
\subfigure[]{
\includegraphics[width=0.45\textwidth, height=0.25\textheight]{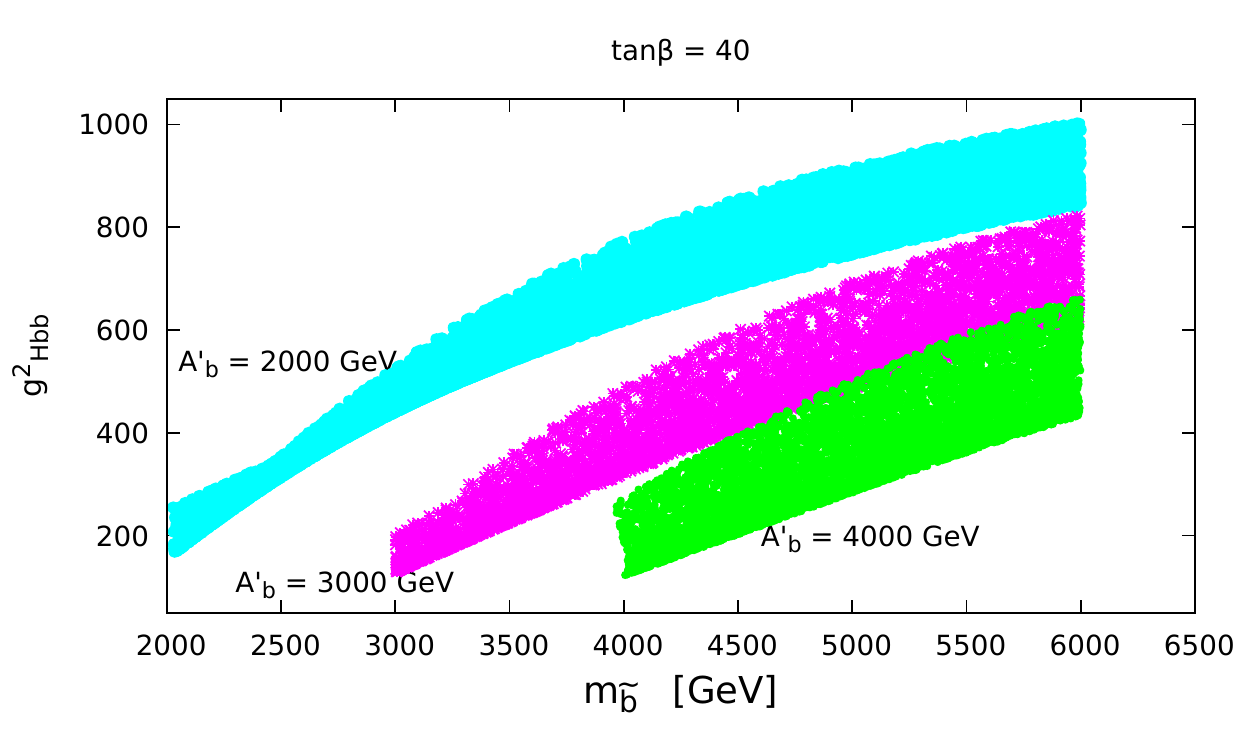}
\label{subfig:msbot-gsq-pabp}
}
\caption{
Variations of $g^2_{Hbb}$ as functions of $\msb$ = $\msbleft$ = $\msbright$ and for three fixed values of $\abprime$ for each case of (a) $\abprime <0$ and (b) $\abprime >0$.
}
\label{fig:msbot-gsq-abp}
\end{figure}
%

As for how the variations of these couplings affect the LHC studies in context, it 
may be recalled that the exclusion region in the $m_A-\tanb$ plane from the 
searches of neutral heavy Higgs bosons \cite{ATLAS:2018gfm, Sirunyan:2018zut, Aad:2020zxo, Aaboud:2017sjh} is shaped by the variations
of the production cross section $\sigma_{(pp \to \bbbar H/A)}$ and the branching fraction BR[$H/A \to \tautaubar$] as functions of $m_A$ and $\tanb$. As far as NHSSM-specific input parameters are 
concerned, while the former is controlled by $\abprime$, the latter are governed 
collectively by $\abprime$ and $\atauprime$. Thus, depending upon the combinations of 
values of $\abprime$ and $\atauprime$, the reported region of exclusion in the
$m_A-\tanb$ plane in the MSSM scenario would get altered (either squeezed or 
extended) when the NHSSM scenario is in context. Quantitatively, however, one can 
perhaps foresee that the largeness of $\Delta_b$ (thanks to the SUSY-QCD correction 
which is missing in $\Delta_\tau$) would control the proceedings
except when the input 
parameters get to be tuned to the contrary. We discuss the decays of the `$H$' and 
the `$A$' states in section \ref{subsec:decays} and their productions in association 
with a $\bbbar$ pair in section \ref{subsec:asso-prod}.

Before we leave this subsection, it would be in order to have a brief discussion on how the other relevant input SUSY 
parameters, viz., $\mgluino$ and $\msbonetwo$ could affect the phenomenology. These parameters enter the proceedings effectively through 
the dominant $\alphas$-dependent term in $\Delta_b$ (see equation 
\ref{eqn:deltab}). In figure \ref{fig:msbot-gsq-abp} we present the variations 
of $g^2_{\Phi bb}$ as a function of $\msb=\msbone=\msbtwo$ for some fixed 
negative (figure \ref{subfig:msbot-gsq-nabp}) and positive (figure 
\ref{subfig:msbot-gsq-pabp}) values of $\abprime$, three of them in each case. For each chosen value of
$\abprime$ we ensure $\msb \geq |\abprime|$ such that appearance of CCB minima could be broadly avoided. Bands for different values of $\abprime$ appear due to a varying $\mgluino$ ($3 \, \mathrm{TeV} \leq \mgluino \, \leq 6 \, \mathrm{TeV}$). The top (bottom) edge of each band corresponds to the largest (smallest) value of $\mgluino$ used in the scan when $\abprime < 0$. The situation is reversed for $\abprime > 0$. The following observations can be made from the plots in figure \ref{fig:msbot-gsq-abp}. 
\begin{itemize}
\item $g^2_{\Phi bb}$ decreases (increases) with increasing $\msb$ for
$\abprime <0 \, (>0)$.
\item $g^2_{\Phi bb}$ increases (decreases) with increasing $\mgluino$ for
$\abprime <0 \, (>0)$.
\item The smallest value ($\approx 1300$) obtained for $g^2_{\Phi bb}$ with
$\abprime <0$ is larger than the largest value ($\approx 1000$) found for the 
same with $\abprime >0$. 
\end{itemize}
Thus, in our present scheme of analysis, larger values of $g^2_{\Phi bb}$, 
for $\abprime <0$, are obtained for a larger $|\abprime| = \msb$ and for a larger
$\mgluino$. The possible maximum for $g^2_{\Phi bb}$, however, tends to 
saturate for larger values of $|\abprime| = \msb$.
On the other hand, for
$\abprime >0$, where we ask how small $g^2_{\Phi bb}$ could get, we find that 
its smallest possible value $\sim 100$ 
(see figure \ref{subfig:msbot-gsq-pabp}) is approached again for
$|\abprime| = \msb$ and for the largest $\mgluino$ that we choose. On this 
occasion, this value of $g^2_{\Phi bb}$ is found to be nearly independent of
$\abprime$. These findings would come in handy when we discuss in section
\ref{subsec:exclusion} how the LHC constraints on the $m_A-\tanb$ plane 
obtained in an MSSM scenario could get altered in the NHSSM scenario.
%
\subsection{Decays of the heavier neutral Higgs bosons}
\label{subsec:decays}
%
In the decoupling regime, the couplings of the heavier neutral Higgs bosons to gauge bosons
($V \equiv W^\pm, Z$), i.e., $g_{HVV}  \propto \cos(\beta-\alpha) \to 0$. This results in a generic insensitivity in their searches in the modes $H \to VV$, $A \to hZ$ etc. Under the circumstances, $H,A \to \ttbar$ would dominate when these Higgs states are sufficiently heavy and $\tanb$ is not large. With increasing $\tanb$ and hence enhanced $y_{b,\tau}$, the decays $H,A \to \bbbar, \tautaubar$ emerge as the only possibilities unless some lighter SUSY states (in particular, the relatively light electroweakinos) are present in the spectrum to which `$H$' and `$A$' could also decay to with moderate branching fractions, a feature that is already present in the MSSM. However, we aim for a simpler setup suited for our purpose where such decays are kinematically forbidden. This is achieved for all $m_{H,A} \leq 3$ TeV by setting $\muprime=1.5$ TeV, $M_{1,2}=2$ TeV and $M_3=3$ TeV (see table \ref{tab:inputs}).

In the NHSSM, decays of $H/A$ to final states involving $b,\tau$ get 
further affected if $\abprime, \atauprime \neq 0$ since these control the magnitudes 
of $\yb$ and $\ytau$, respectively. As expected, this could well modify the 
experimental sensitivities of searches of heavier Higgs bosons in these final states. 
Consequently, the exclusion region in the $m_A-\tanb$ plane as obtained for the MSSM 
scenario would get altered for the NHSSM case. Here, it is important to point out 
that although the decays of heavier Higgs bosons to bottom quarks dominate at larger 
$\tanb$ (and large negative $\abprime$), final states involving $b$-jets are prone to 
large jetty backgrounds from the SM. Hence searches for the heavy Higgs states at the 
LHC are best carried out in the subdominant modes involving the $\tau$'s. Note, 
however, that while $y_b$ (and not $\ytau$) controls $\sigma(pp \to \bbbar H/A)$,
BR[$H/A \to \tautaubar$] are controlled by both
$\yb$ and $\ytau$. The highest sensitivity to the $\tautaubar$ final state is thus attained when the right balance in the magnitudes of $\yb$ and $\ytau$ is struck over the NHSSM parameter space.

Given that the magnitudes and the patterns of variations of branching fractions to
$\bbbar$ and $\tautaubar$ of the `$H$' and `$A$' states are expected (and found) to be very similar, a common representative illustration of these variations for the two states would suffice. In addition, the complementarity of BR[$H/A \to \bbbar$] and BR[$H/A \to \tautaubar$] allows us to discuss the proceedings in terms of any one of these quantities. The obvious choice is then BR[$H/A \to \tautaubar$] since the
concerned experimental analyses consider the decays $H/A \to \tautaubar$.
It may further be noted that BR[$H/A \to \tautaubar$] depends not only on $\tanb$ and $\atauprime$ but also on $\abprime$. This is since $\abprime$ influences the partial decay width for $H/A \to \bbbar$ which, in turn, affects BR[$H/A \to \tautaubar$]
by altering the total decay width of $H/A$ that enters the computation of the branching fractions.
%
\begin{figure}[t]
\centering
\subfigure[\label{subfig:br-tautau-tb10}]{
\includegraphics[width=0.38\textwidth,height=0.23\textheight]{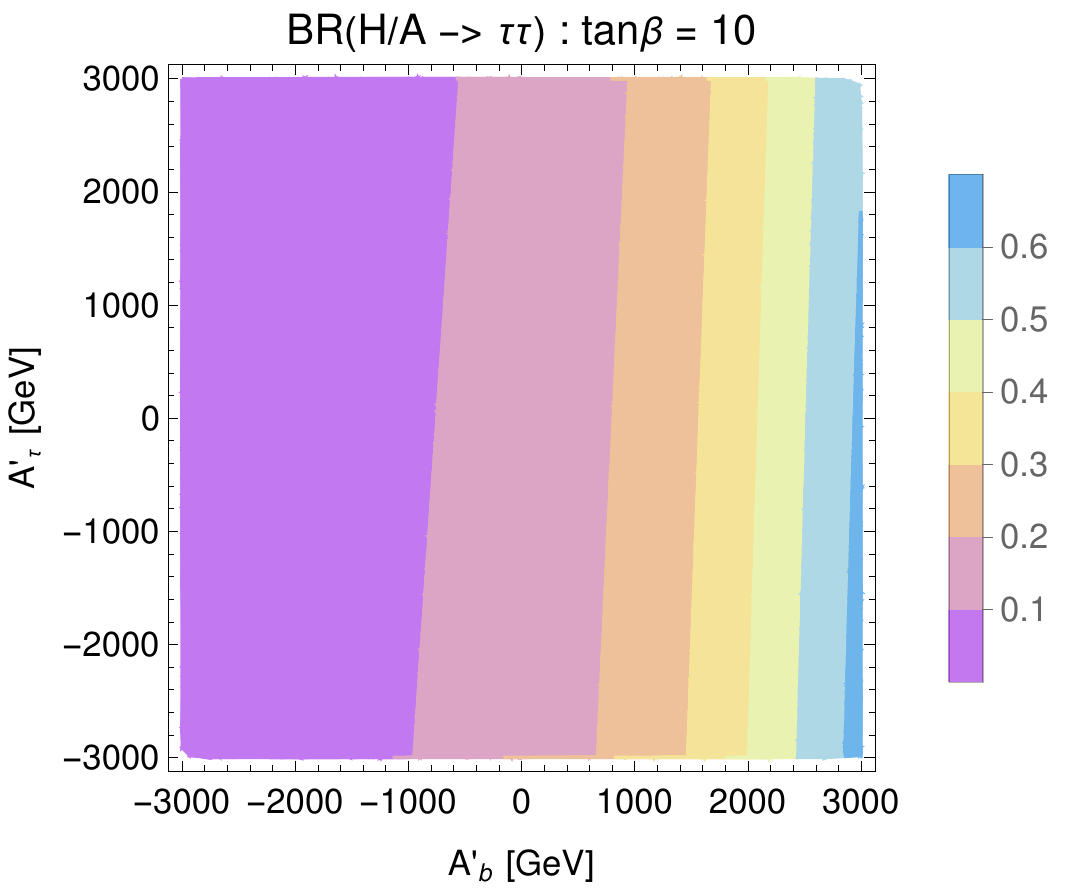}}
\hskip 25pt
\subfigure[\label{subfig:br-tautau-tb50}]{
\includegraphics[width=0.38\textwidth,height=0.23\textheight]{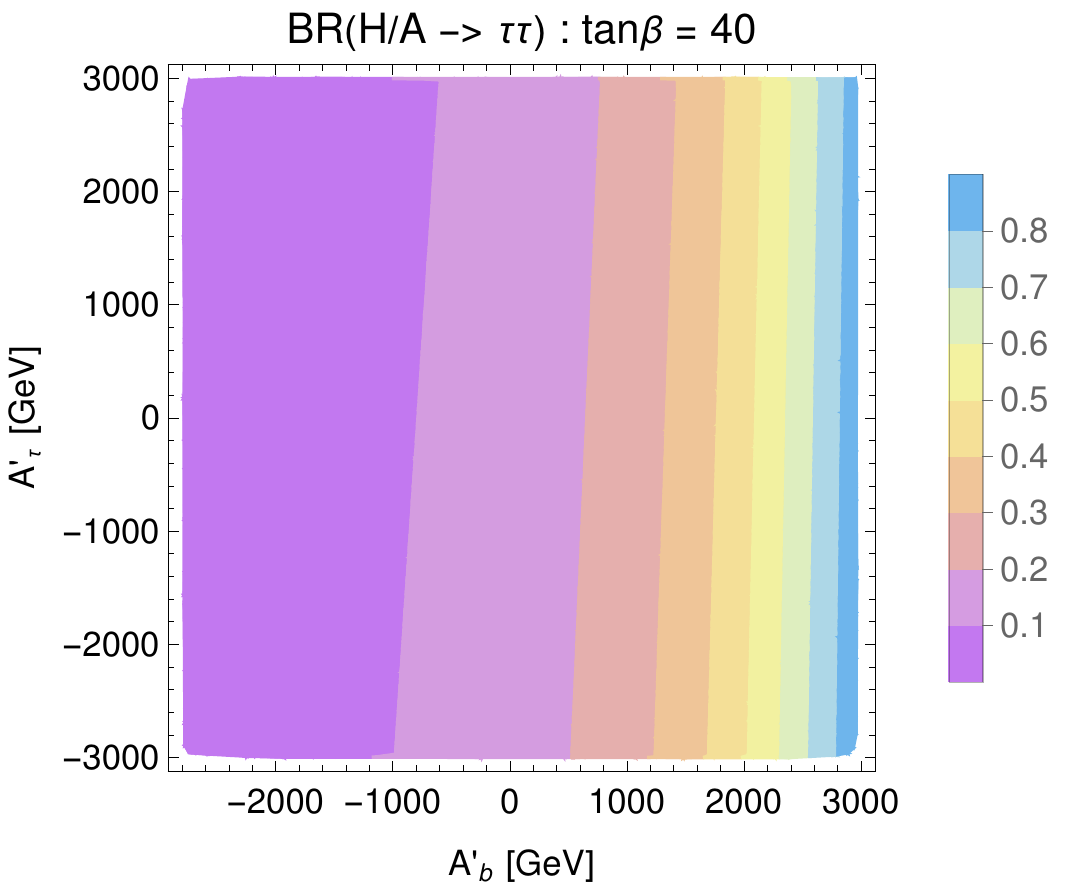}}
\caption{Variations of BR[$H/A \to \tautaubar$] in the $\abprime-\atauprime$ plane for (a) $\tanb=10$ and (b) $\tanb=40$ and for $m_{H,A}=2.5$ TeV.
}
\label{fig:br-tautau-abp-ataup}
\end{figure}
%

In figure \ref{fig:br-tautau-abp-ataup} we present the variations of
BR[$H/A \to \tautaubar$] in the $\abprime-\atauprime$ plane for $\tanb=10$ (figure \ref{subfig:br-tautau-tb10}) and $\tanb=40$ (figure \ref{subfig:br-tautau-tb50}) when $m_{H,A}$ are set at a value of 2.5 TeV.
It is seen that, irrespective of the value of
$\tanb$, the dependence of BR[$H/A \to \tautaubar$] on $\atauprime$ is only rather mild. Thus, it is $\abprime$ which almost single-handedly dictates
BR[$H/A \to \tautaubar$]. As discussed in section \ref{subsec:higgs-sector}, this can be traced back to the smallness of $\Delta_\tau$ when compared to $\Delta_b$ as the former lacks the gluonic (strong) contribution. It may further be noted how rapidly
BR[$H/A \to \tautaubar$] drops as $\abprime$ decreases from its large positive value. Hence, for any given $\tanb$, the only situation when decays to $\tautaubar$ could be 
comparable to (or may even dominate) those to $\bbbar$ is when $\abprime$ has a large 
enough positive value thus rendering the couplings $H/A \bbbar$ much smaller compared 
to those for $H/A \tautaubar$. As can be found from figure
\ref{fig:br-tautau-abp-ataup}, under such a circumstance, BR[$H,A \to \tautaubar$] 
could reach $\sim 90$\% when $\tanb$ is large while BR[$H,A \to \tautaubar$] dropping to a mere
$\sim 10$\%. 

It is important here to note that, for any given integrated luminosity, for the LHC 
experiments to be sensitive to searches for the $H/A$ states in the mode
$pp\to \bbbar H,A$ followed by $H,A \to \tautaubar$, the product
$\sigma(pp \to \bbbar H/A) \, \times$ BR[$H/A \to \tautaubar$] has to be optimally 
large. With $\abprime$ effectively controlling both these quantities but that being 
in a contrasting manner, only an optimal value of $\abprime$ would maximize the said 
product, for any given set of other input parameters.
%
\subsection{Production of heavier neutral Higgs bosons ($H,A$) with a bottom quark pair}
\label{subsec:asso-prod}
%
Predicting the production cross section of a Higgs boson in association with a pair 
of $b$-quarks at hadron colliders involves theoretical subtleties. $m_b$ being small 
yet may not be ignored, there are two competing approaches to calculate the hard, 
parton level cross section: the so-called 4-flavor scheme (4FS) and the 5-flavor 
scheme (5FS). These entail different perturbation expansions of the same physical 
observables, i.e., $pp \to \mathrm{Higgs} + jets$, which start contributing at 
different perturbation orders in $\alpha_s$. Their close agreement is then expected 
only at sufficiently high orders \cite{Dawson:2005vi, Campbell:2004pu}.

\begin{figure}[t]
\begin{center}
\framebox{\cred{4FS LO}} \hskip 5pt
\subfigure[]{
\includegraphics[width=0.22\textwidth,height=0.11\textheight]{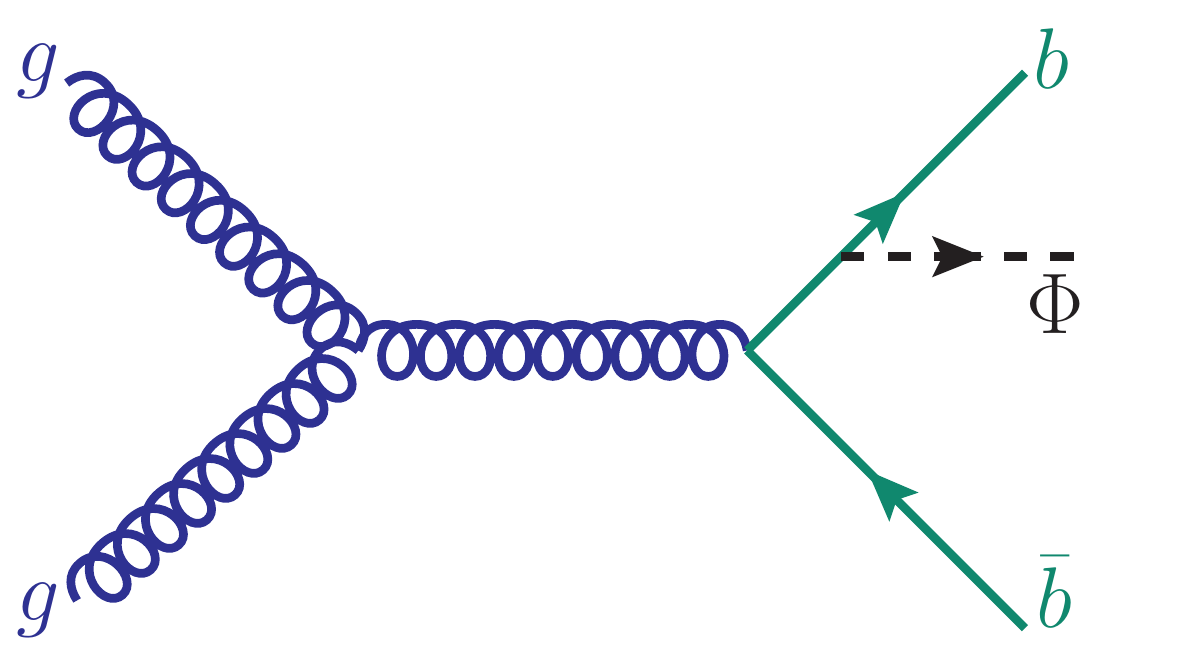}
\label{subfig:gg-4fs-schan}
}
\hskip 25pt
\subfigure[]{
\includegraphics[width=0.22\textwidth,height=0.11\textheight]{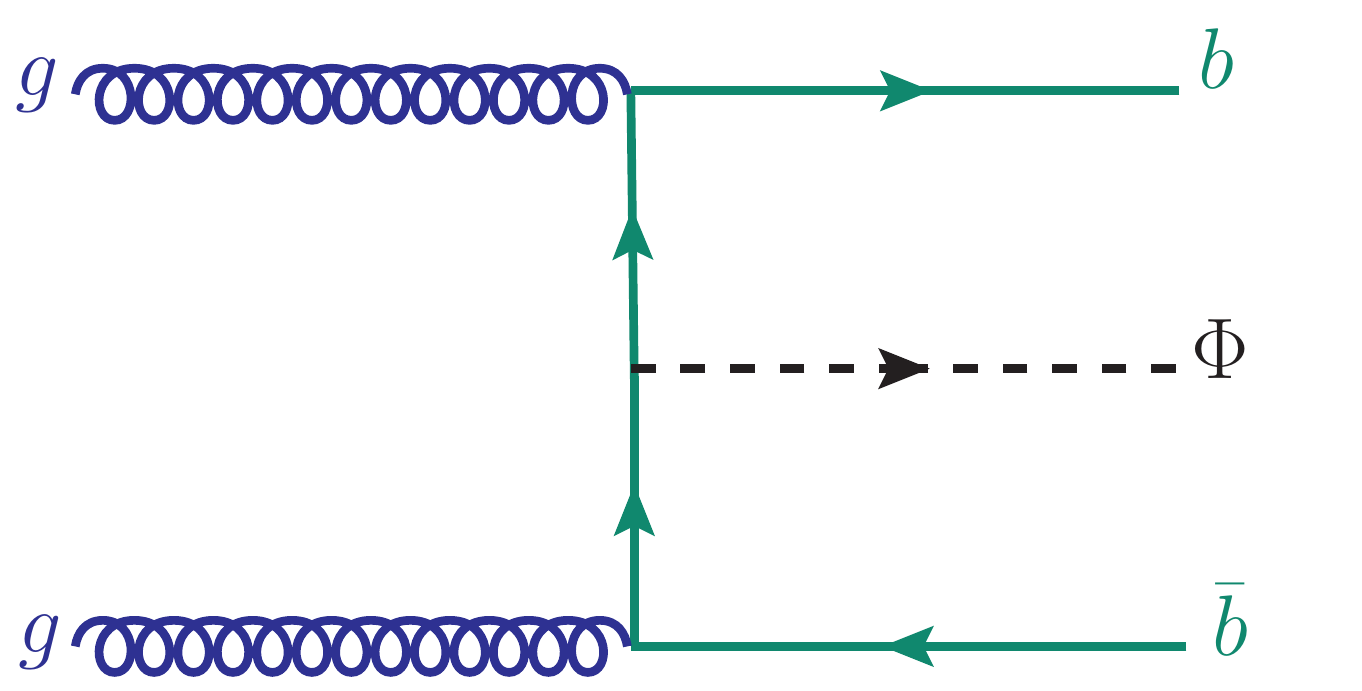}
\label{subfig:gg-4fs-tchan}
}
\hskip 25pt
\subfigure[]{
\includegraphics[width=0.22\textwidth,height=0.11\textheight]{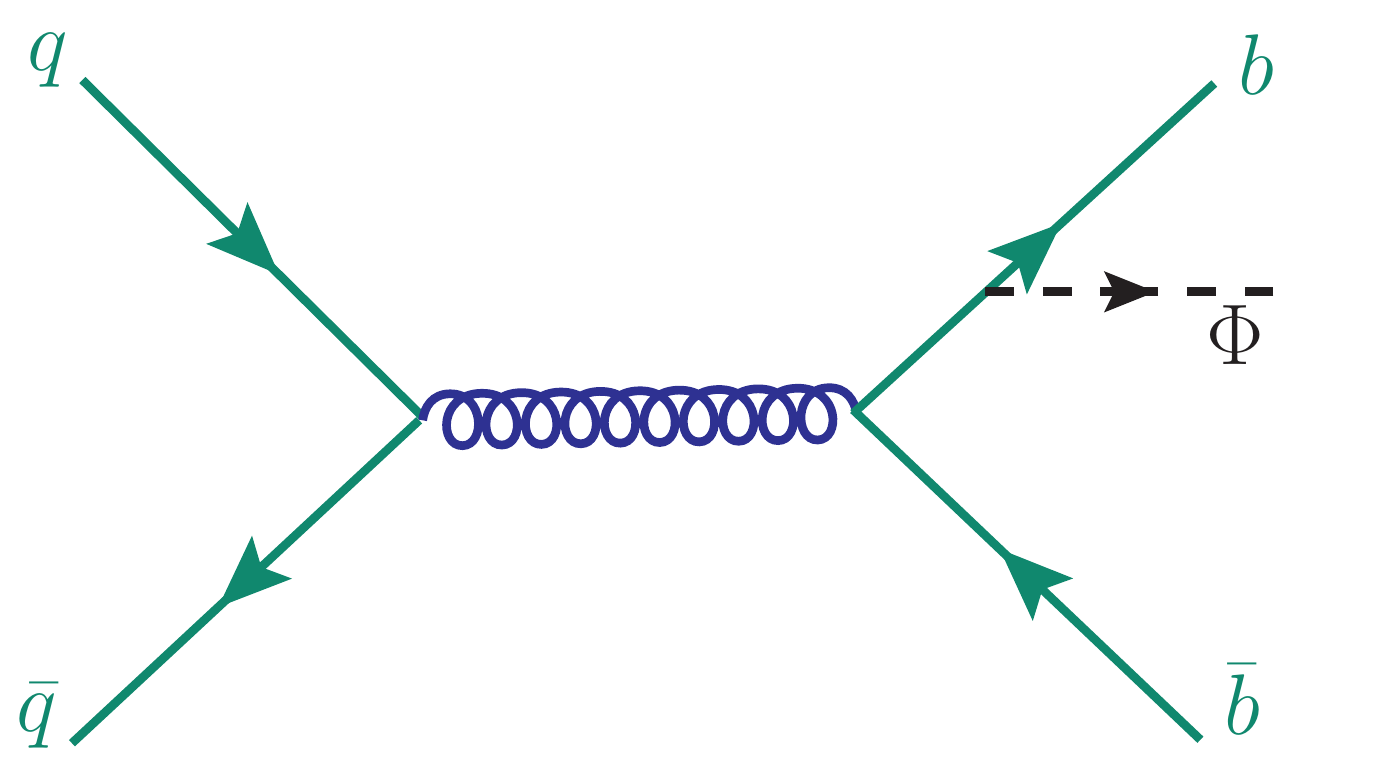}
\label{subfig:qqbar-4fs}
} \\
\vskip 25pt
\hskip 20pt
\framebox{\cred{5FS}} \hskip 5pt
\subfigure[]{
\includegraphics[width=0.22\textwidth,height=0.12\textheight]{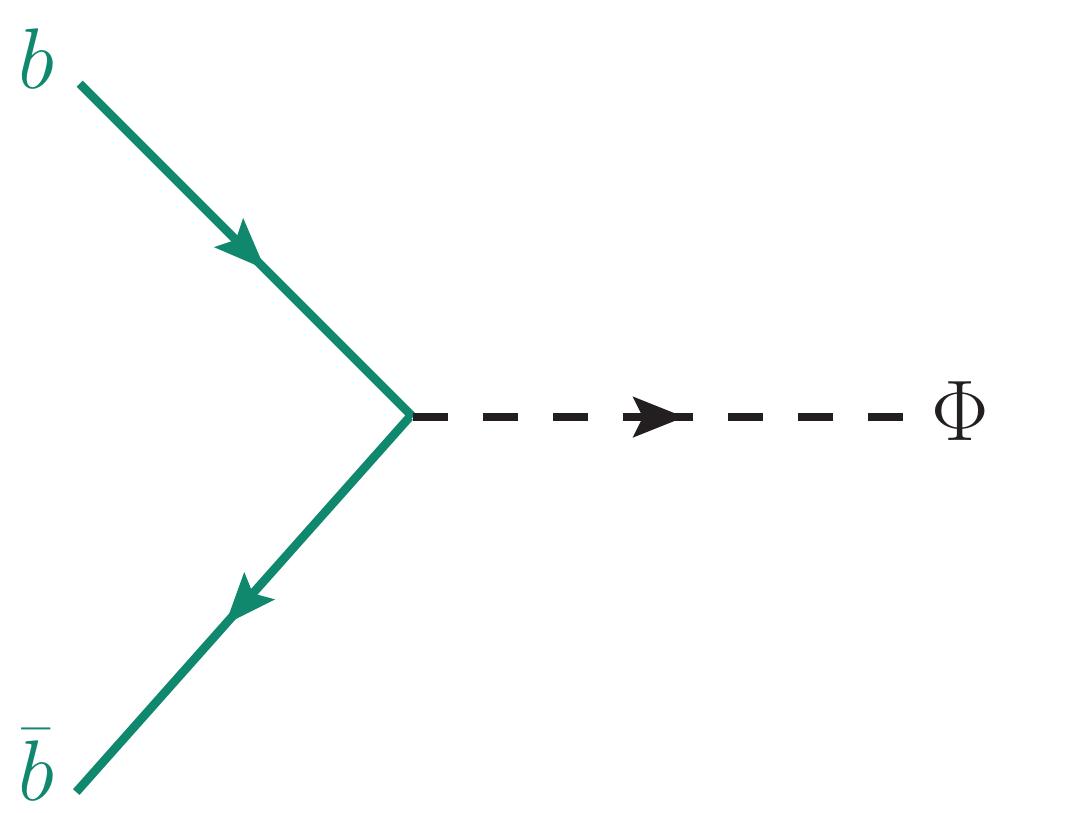}
\label{subfig:bbbar-phi-5fs}
}
\hskip 25pt
\subfigure[]{
\includegraphics[width=0.22\textwidth,height=0.11\textheight]{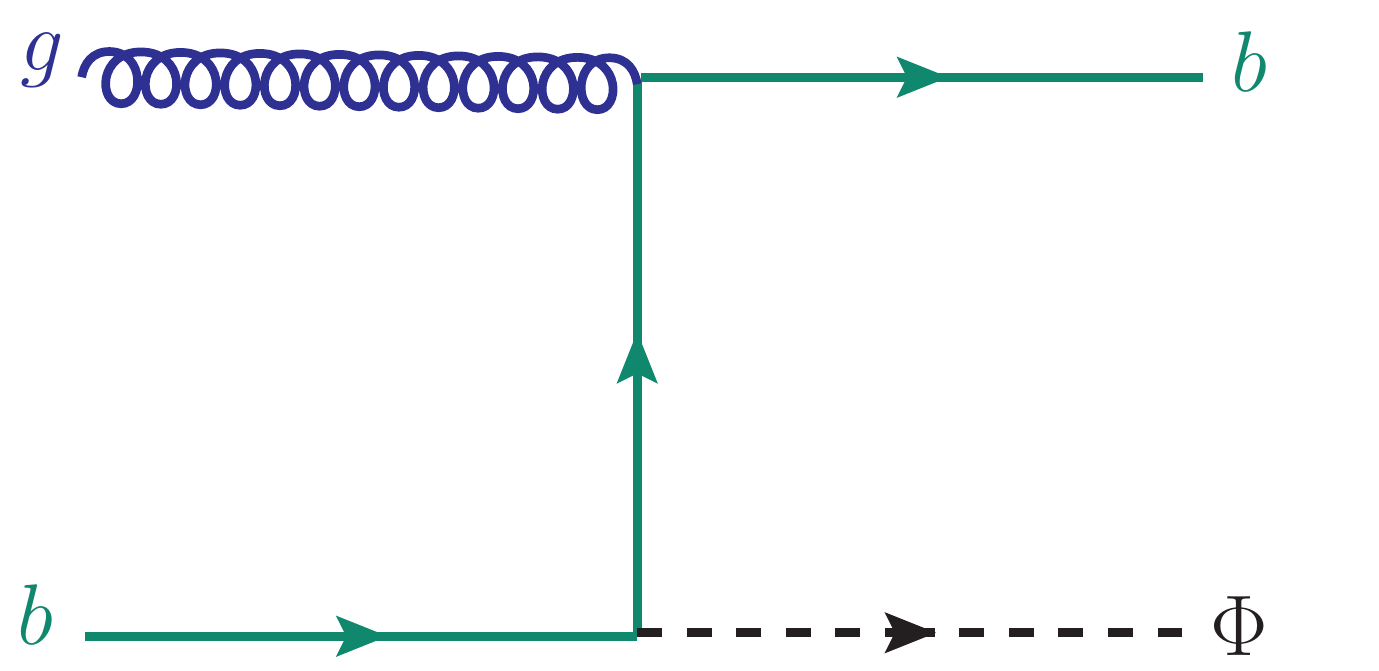}
\label{subfig:gb-5fs-tchan}
}
\hskip 25pt
\subfigure[]{
\includegraphics[width=0.22\textwidth,height=0.12\textheight]{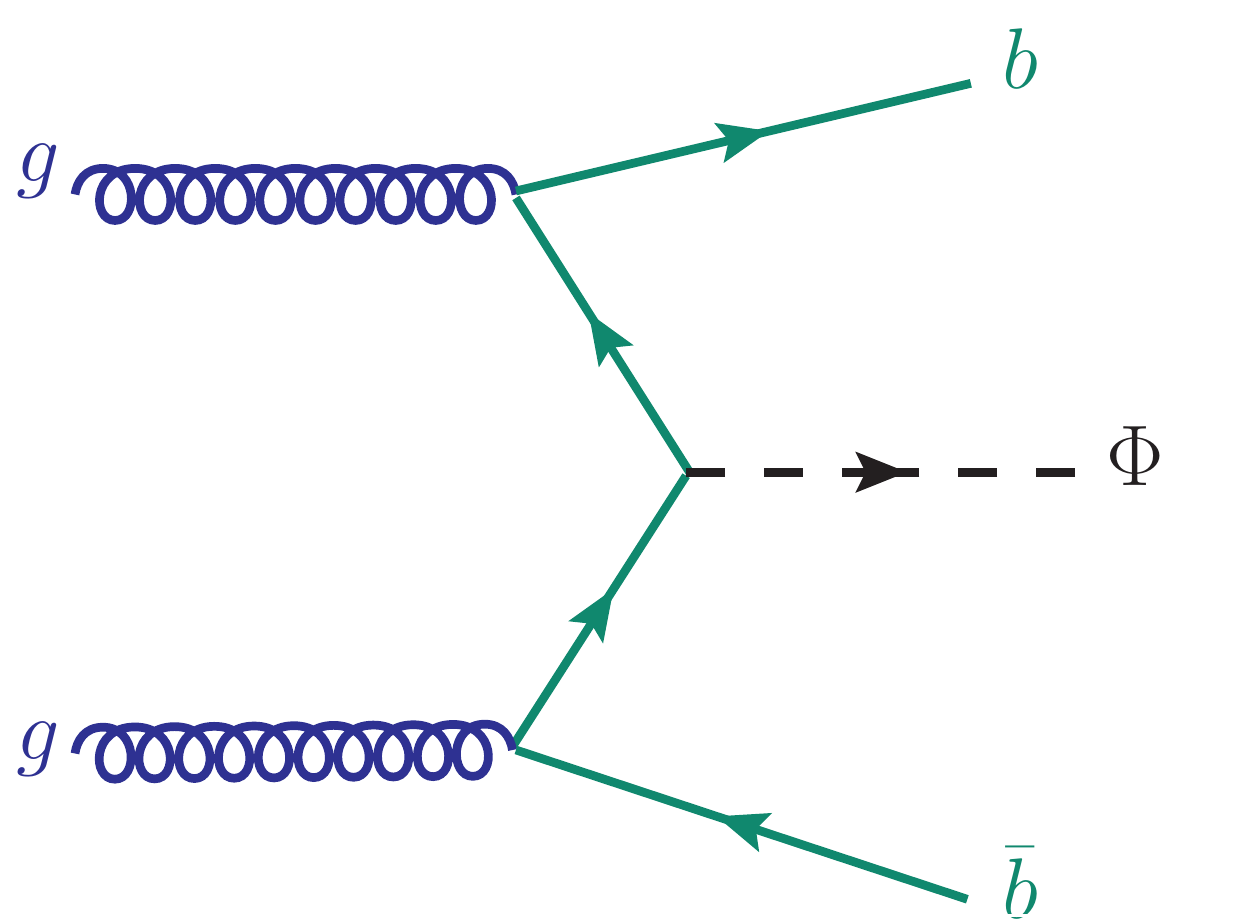}
\label{subfig:gg-5fs-tchan}
}
\caption{Representative 4FS LO Feynman diagrams contributing to exclusive
$b\bar{b}\Phi$ ($\Phi= h,H,A$) production in the 4FS (top panel) and the ones for the 
corresponding inclusive process at LO (left), NLO (middle) and NNLO (right) in the 
5FS (bottom panel). See text for details.}
\label{fig:feynman-diagrams}
\end{center}
\end{figure}

In the 4FS, the $b$-quark is taken to be massive. Hence the proton (or the 
antiproton)  parton distribution contains gluons and the four light
($\sim$massless) flavor quarks, $u,d,s,c$ and the corresponding antiquarks. The 
relevant parton level processes at the LHC, at the lowest order (LO) in 
perturbation theory, are then driven by gluon-fusion and quark-antiquark 
annihilation, i.e.,
\beq
gg, q\bar{q} \to \bbbar~ H/A \;.
\eeq
A generic, representative set of Feynman diagrams for this (4FS LO) category is shown 
in the upper panel of figure \ref{fig:feynman-diagrams}. At the LHC energies, the LO 
cross section is largely dominated by the gluon fusion process. Note that the
$t$-channel gluon-fusion process in figure \ref{subfig:gg-4fs-tchan} leads to large 
logarithms $\sim \ln(\mu_F^2/m_b^2)$ on integrating over the phase space of the final 
state $b$-quarks; $\mu_F \sim m_{_{\Phi \in H,A}}$ being the so-called factorization 
scale. In the computation of an inclusive cross section of the sort, such large 
logarithms (involving finite $m_b$) are effectively resummed to all orders via the 
introduction of the $b$-quark parton distribution function (PDF)
\cite{Barnett:1987jw} in the proton with $m_b \to 0$. This is the 5FS. A representative set of Feynman diagrams
in this scheme, for the inclusive process in reference, is shown in the lower panel of figure \ref{fig:feynman-diagrams}.
%
%

The fully inclusive cross section in the 5FS would then be dominated by the LO 
process $\bbbar \to \Phi$ \cite{Dicus:1988cx} in figure \ref{subfig:bbbar-phi-5fs} 
and exceeds the corresponding 4FS LO estimates. This can be understood by noting that 
the dominant 4FS LO process (the $t$-channel one of figure \ref{subfig:gg-4fs-tchan}) 
appears only at the next-to-next-to-leading order (NNLO)-QCD in the 5FS, as shown in
\ref{subfig:gg-5fs-tchan}. It is further observed that the 4FS (5FS) approach is 
better suited in the asymptotic limit of small (large) Higgs boson masses when the 
mass of the bottom quark cannot (can) be neglected as is the defining characteristic 
of the scheme \cite{Harlander:2011aa}.

Experiments studying $pp \to \bbbar \Phi$ necessarily have to tag $b$-jet(s). An 
exclusive search like this requires a minimum $p_T^{(b)}$ which ensures that the tagged
$b$-jets do not have their origins in the $b$-quark PDF. Hence one may rely 
exclusively on the 4FS for the LO rates. This further guarantees that the 
Higgs boson to have been radiated off one such final state $b$-quark produced in the hard scattering thus probing $y_b$ directly.

Eventually, bounds derived from the experiments on $\sigma_{(pp \to bb H/A)} \times 
\mathrm{BR}[H/A \to \tautaubar])$, and on the $m_A-\tanb$ plane thereon
\cite{ATLAS:2017eiz, CMS:2018rmh, ATLAS:2020zms}, refer to the inclusive (total) 
cross sections. In either of these two schemes, more precise estimates of these cross 
sections, which are stable against variations in the unphysical
factorization/renormalization ($\mu_R$) scales, can only be found by including higher order corrections. 

Given that the computation of the higher-order corrections is simpler in the 5FS
(since $m_b \rightarrow 0$ and the LO process is a resonant one), cross sections
up to NNLO (in $\alphas$) are known there for long
\cite{Dicus:1998hs, Balazs:1998sb, Harlander:2003ai}. In contrast, rates only up to
next-to-leading order (NLO) are available in the 4FS
\cite{Dittmaier:2003ej, Dawson:2003kb, Wiesemann:2014ioa}.\footnote{This calculation is recently improved to contain NLO-electroweak corrections \cite{Pagani:2020rsg}.} 
At these available orders, the cross sections obtained in the 
two schemes tend to show improved agreements
\cite{Campbell:2004pu} along with a much milder dependence on $\mu_F$ and $\mu_R$.
In spite of this, 4FS NLO results are unable to capture logarithmic terms beyond the first few while the 5FS NNLO ones fail to do so for the so-called power-suppressed terms \cite{Wiesemann:2014ioa}.\footnote{Of recent, next-to-next-to-next-to leading order ($\mathrm{N^3LO}$) \cite{Duhr:2019kwi} and N$^3$-leading log (N$^3$LL) \cite{Ajjath:2019neu} estimates in the 5FS have become available and are found to agree even more closely with the NLO-QCD one in the
4FS.}

Capturing the best from the two different kinematic regimes requires a systematic combination of these contributions \cite{Harlander:2011aa, Aivazis:1993pi, Thorne:1997ga, Cacciari:1998it, Kramer:2000hn, Tung:2001mv, Thorne:2006qt}. Here, we adopt  an improved approach called FNOLL-B \cite{Forte:2015hba, Bonvini:2015pxa, Bonvini:2016fgf, Forte:2016sja} as also done by a recent ATLAS analysis \cite{ATLAS:2020zms}. This approach combines the 4FS-NLO accuracy with the accuracy of NNLO leading-log (NNLL) for the resummed collinear logarithms of the 5FS.\footnote{Recently, an even more accurate prediction of the inclusive production cross section by matching 4FS-NLO results with those from the 5FS N$^3$LO ones has been made \cite{Duhr:2020kzd}.}
Towards this, we follow the recommendations of references \cite{LHCHiggsCrossSectionWorkingGroup:2016ypw, recommend}. Hence we use the cross sections for $pp \to \bbbar H$ quoted there as a function of $m_H$ \cite{Bonvini:2015pxa, Bonvini:2016fgf} where the $Hbb$ coupling is taken to be SM-like and 
the common factorization ($\mu_F$) and the renormalization ($\mu_R$) scales, denoted by $\mu_{_{FR}}=\mu_F=\mu_R$, is set to
$(m_H+2m_b)/4$.\footnote{Note that $\mu_{_{FR}}$ is much smaller than $m_H$ itself and is known to lead to reasonable agreements between 4FS and 5FS estimates already at LO and NLO \cite{Harlander:2003ai, Maltoni:2003pn, Boos:2003yi} with improved perturbative convergence \cite{Bonvini:2015pxa, Maltoni:2003pn, Maltoni:2005wd, Maltoni:2012pa, Harlander:2015xur}.}
These rates are indicated to be cross-checked against the FNOLL-B matched results of reference \cite{Forte:2016sja}.

We obtain $\sigma(pp \to \bbbar H/A)|_{\mathrm{\mbox{\tiny NHSSM}}}$ by an appropriate 
rescaling of $y_b$ \cite{Dittmaier:2006cz, Dawson:2011pe, Dittmaier:2014sva}, i.e., 
by multiplying $\sigma(pp \to \bbbar \hsm)$ by the square of the factors
$g_{\bbbar H/A}$ of equation \ref{eqn:couplings} that are now computed for the 
NHSSM. It should be noted that, over the broad range of $m_{H/A}$ we consider, their
mass-differences are found to be smaller than the experimental resolution
(except when $m_A \leq 400$ GeV). Hence these states are treated as
mass-degenerate \cite{ATLAS:2020zms}. Thus, for any given mass $m_{H/A}$, their contributions are just added up.

\begin{figure}[!htb]
\begin{center} 
\subfigure[]{%
\includegraphics[width=0.32\textwidth,height=0.20\textheight]{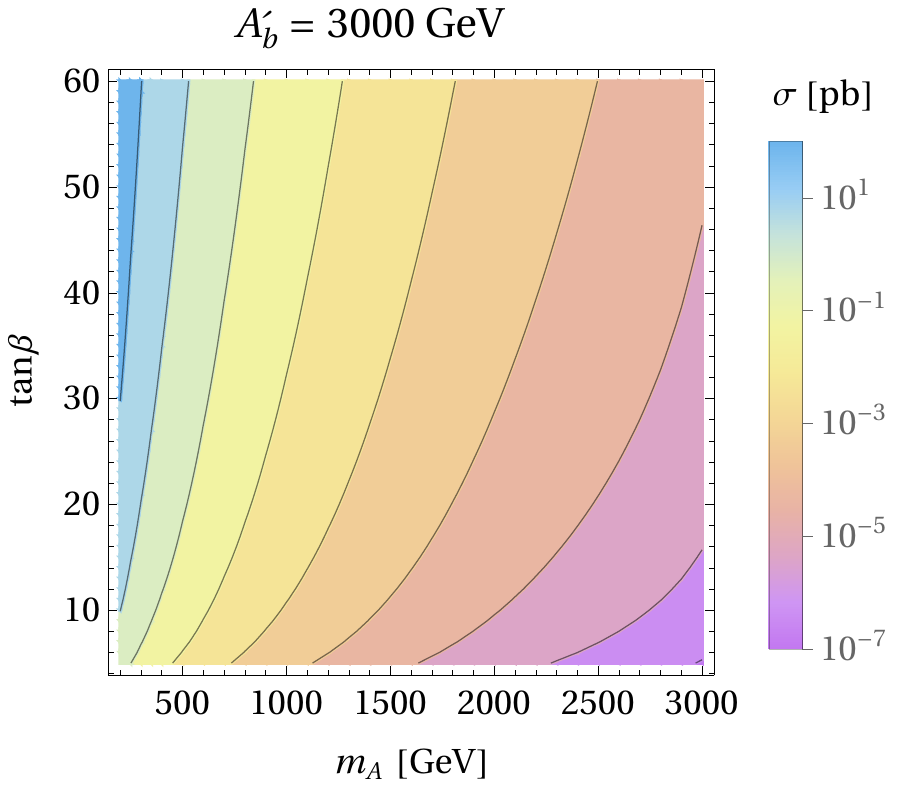}
\label{subfig:xsec-ma-tanb-abp3}
}
\subfigure[]{%
\includegraphics[width=0.32\textwidth,height=0.20\textheight]{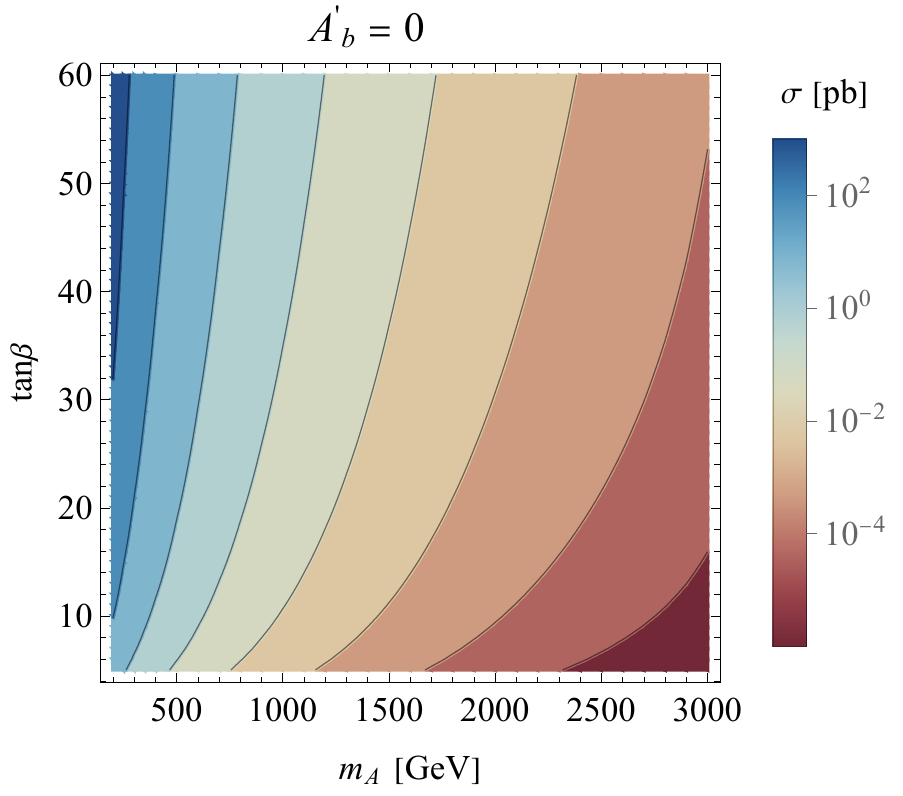}
\label{subfig:xsec-ma-tanb-abp0}
}
\subfigure[]{%
\includegraphics[width=0.32\textwidth,height=0.20\textheight]{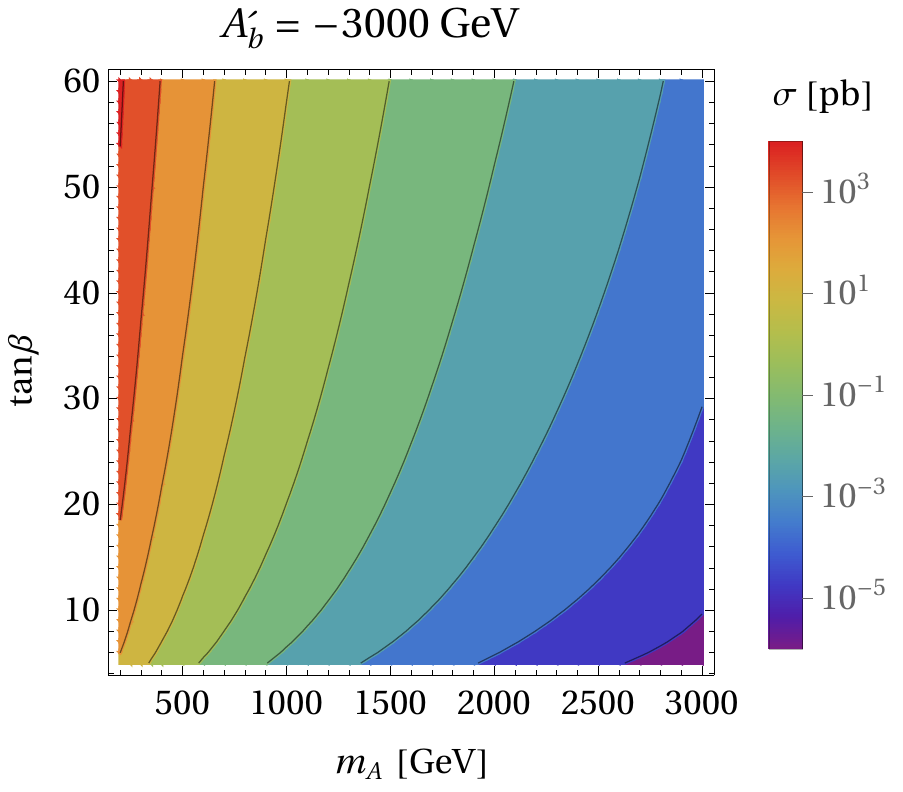}
\label{subfig:xsec-ma-tanb-abp-3}
}
\caption{Variations of cross sections $\sigma_{(pp \to \bbbar H + \bbbar A)}$ (in pb) at the 
13 TeV LHC in the $m_A-\tanb$ plane for (a) $\abprime =3$ TeV, (b) $\abprime=0$ and 
(c) $\abprime= -3$ TeV.
}
\label{fig:xsec-ma-tanb}
\end{center}
\end{figure}

In figure \ref{fig:xsec-ma-tanb} we present the variations of the combined cross sections for $pp\to \bbbar H,A$ at the 13 TeV LHC in the $m_A$--$\tanb$ plane for
(a) $\abprime =3$ TeV, (b) $\abprime=0$ (the MSSM-like case) and (c) $\abprime= -3$ 
TeV. As can be seen from these plots, for large $\tanb$ and larger $m_A$, the 
cross sections with $\abprime =-3$ TeV could be up to a couple of orders larger 
in magnitude than that when $\abprime =3$ TeV. The plot in the middle ($\abprime=0$) serves closely as an MSSM-reference for the purpose in which, for any given values of $m_A$ and $\tanb$, one finds intermediate values of cross sections when compared to the former two cases. The patterns of variations of the cross section 
are intimately related to the same for the respective variation of $y_b$ (see figure 
\ref{fig:yb-tanb-abprime}), or for that matter, of $g^2_{\bbbar H/A}$ (see figure 
\ref{fig:gffsqphi}) in the same parameter-plane and these shape the projected exclusion regions of the parameter space of the NHSSM scenario as derived in section \ref{subsec:exclusion}.
%
%
\subsection{Constraining $m_A-\tanb$ plane: NHSSM vis-a-vis LHC exclusion (in the MSSM)}
\label{subsec:exclusion}
%
As pointed out in section \ref{subsec:bb-tautau-phi}, the sensitivity of a search for 
a heavy Higgs boson via its associated production with a $b$-quark pair, at the LHC, 
broadly depends on the quantity $\sigma(pp \to \bbbar H/A)$ times the branching 
fraction of $H/A$ into an optimally sensitive mode which happens to be the
$\tautaubar$ mode. Recent LHC analyses
\cite{ATLAS:2017eiz, CMS:2018rmh, ATLAS:2020zms} have put constraints (at 95\% confidence level (CL)) on the maximum 
allowed values of the product
$\sigma(pp \to \bbbar \Phi) \times \mathrm{BR}[\Phi \to \tautaubar]$ as a function of 
$m_\Phi$, in a model-independent way. These are then
translated\footnote{This is possible since the narrow-width approximation is broadly reliable as $\Gamma_\Phi \over m_\Phi$ ($\Phi \ni H,A$) remains on the smaller side (varying between 0.5\% -- 5\%) over the region of the NHSSM parameter space chosen for the present study that includes large values of both $\tanb$ and $|A_b'|$.} into a
model-dependent (MSSM) exclusion contour (again at 95\% CL) in the plane of $m_A$ and $\tanb$, the sole 
unknown parameters of the theoretical setup in reference on which the observable 
quantity under consideration depends. Given that both $\sigma(pp \to \bbbar H/A)$ and   
$\mathrm{BR}[H/A \to \tautaubar]$ crucially depend on an NHSSM-specific parameter 
like $\abprime$, in addition to their usual (MSSM-type) dependencies on $m_A$ and
$\tanb$, the existing domain of exclusion in the $m_A-\tanb$ plane for the MSSM-like limit (i.e., $\abprime=0$) is bound to get altered in the NHSSM as $\abprime$ varies. We 
work out such exclusion contours in the NHSSM scenario and contrast those with the one from a recent ATLAS analysis \cite{ATLAS:2020zms} that provides the strongest exclusion till date on the 
$m_A-\tanb$ plane.

\begin{figure}[t]
\begin{center}
\subfigure[]{
\includegraphics[width=0.46\textwidth,height=0.29\textheight]{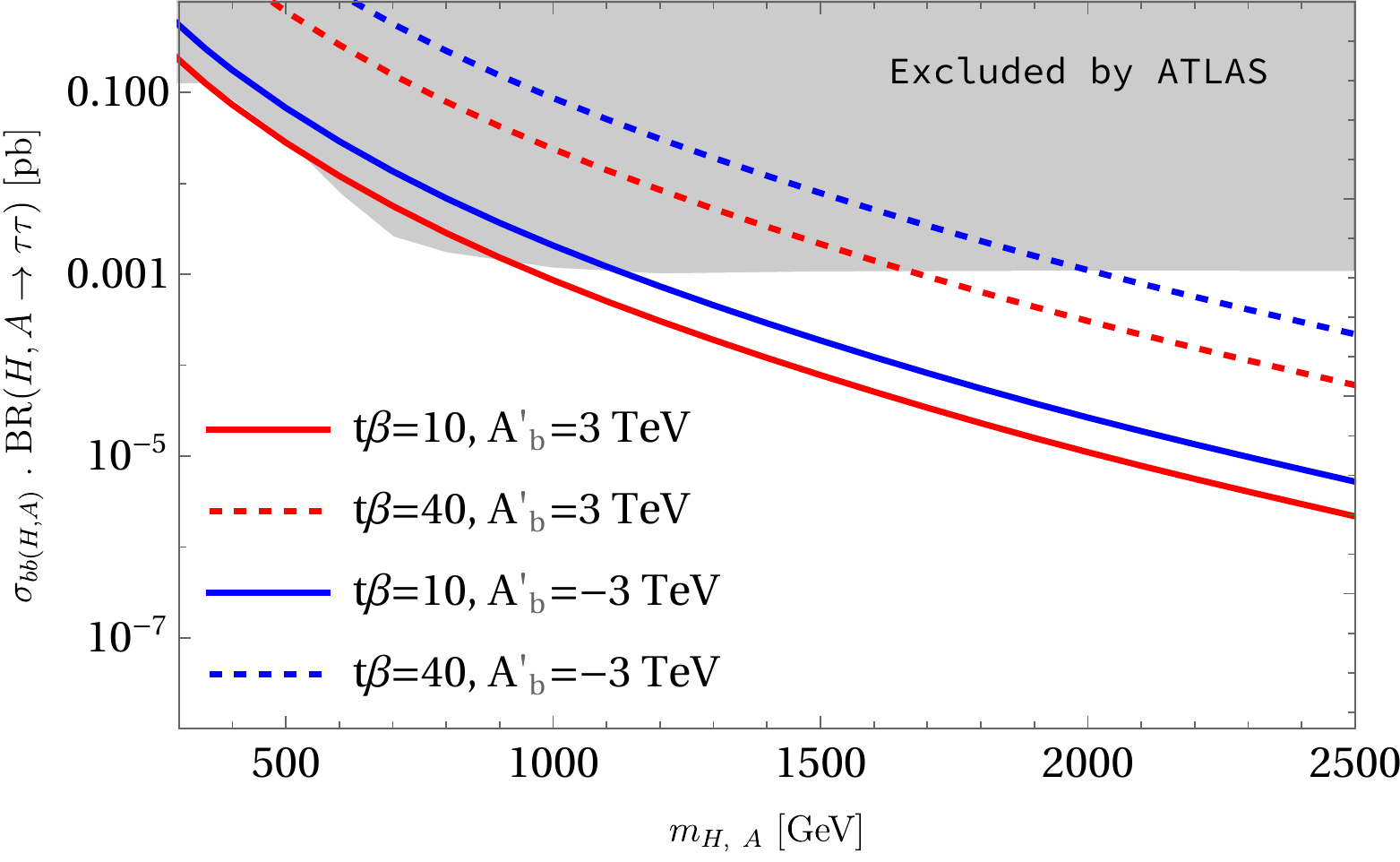}
\label{subfig:sigbr-tb-abp}
}
\hskip 25pt
\subfigure[]{
\includegraphics[width=0.44\textwidth,height=0.30\textheight]{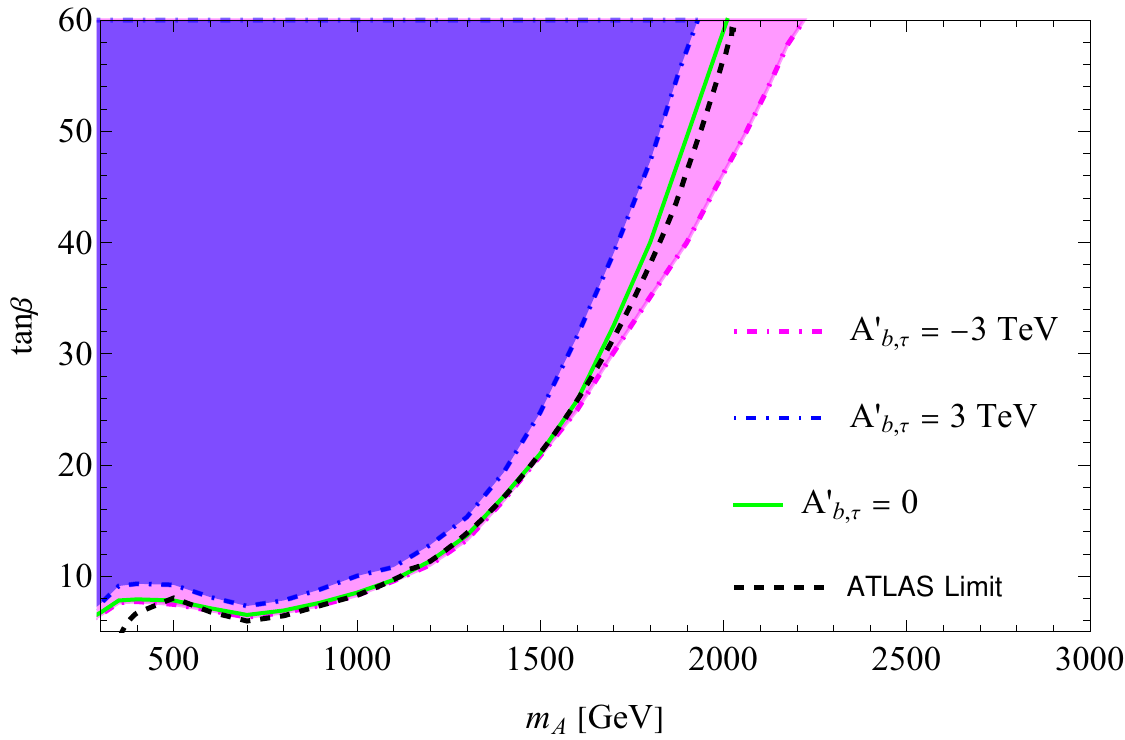}
\label{subfig:ma-tb-excl}
}
\vskip -2pt
\caption{(a) Varying $\sigma \times \mathrm{BR}$ for different values of $\abprime$ and
$\tanb$, as a function of $m_{\Phi=H,A}$, shown against its observed upper limits at 95\% CL as reported in figure 2(b) of reference \cite{ATLAS:2020zms} and defined by the lower/left edge of the grey region and
(b) the results of recasting the observed ATLAS upper limits on $\sigma \times \mathrm{BR}$ in figure \ref{subfig:sigbr-tb-abp} on the $m_A-\tanb$ plane for various representative values of $\abprime$. In figures \ref{subfig:sigbr-tb-abp} and
\ref{subfig:ma-tb-excl}, curves representing the reinforced (relaxed) bounds are obtained using the combination $\abtauprime = -(+) 3$ TeV.
}
\label{fig:exclusions}
\end{center}
\end{figure}

In figure \ref{subfig:sigbr-tb-abp} we study the variations of
$\sigma \times \mathrm{BR}$ as functions of $m_{_{H/A}}$ for two discrete values of each of $\tanb$ (10 and 40) and $\abprime$ ($-3$ TeV and 3 TeV)\footnote{As mentioned earlier, `$H$' and `$A$' states could practically be considered mass-degenerate over the considered range of their masses. Hence, for the purpose in hand, for any given mass $m_A=m_H$, their production cross sections are added up and then multiplied by a branching fraction to $\tautaubar$ which happens to be essentially the same for the two Higgs states,
in particular, in the scenario under consideration where their decays to only the SM states are kinematically allowed.}
and contrast those with its observed upper limits (at 95\% confidence limit (CL)) as reported by the ATLAS collaboration (in figure 2(b) of reference \cite{ATLAS:2020zms}) and represented by the lower/left edge of the grey region 
in figure \ref{subfig:sigbr-tb-abp}. 
The sections of the blue and the red curves falling within the 
grey region indicate the values of $m_{_{H/A}}$ that are ruled out in the NHSSM 
for different combinations of $\tanb$ and $\abprime$. As is expected, the largest 
(smallest) of the values of $m_{_{H,A}}$ is ruled out with the dashed-blue
(solid-red) curves for which $\tanb$ is large (small) and $\abprime$ is negative 
(positive).
%
%

The same analysis for the NHSSM is then ported to delineate the altered excluded regions in the $m_A-\tanb$ plane vis-a-vis the one presented by the mentioned ATLAS experiment \cite{ATLAS:2020zms} for a generic MSSM scenario called the $M_h^{125}$ scenario \cite{Bagnaschi:2018ofa} whose defining features are all preserved in the
NHSSM scenario we adopt in this work. While contrasting our findings with the experimental one, we limit ourselves to the associated $\bbbar H,A$ productions whereas the experiment considers the gluon fusion process as well for the production of these Higgs states. It may, however, be noted that at larger values of $\tanb$, where we find the NHSSM effects to be somewhat prominent, $pp \to \bbbar H,A$ play the dominant role \cite{Aaboud:2017sjh}. Hence, given the scope of the present work, our approach suffices.
The closely overlapping dashed contours in green and black in figure
\ref{subfig:ma-tb-excl} represent our derivation of the exclusion region with
$\abprime=0$ (the MSSM limit for the concerned sector) and the same obtained for the 
$M_{h}^{125}$-MSSM scenario by the ATLAS experiment (at 95\% CL; in figure 2(c) of reference \cite{ATLAS:2020zms}), 
respectively. In NHSSM scenarios with
$\abprime \neq 0$, one finds a slightly relaxed exclusion (to the edge of the purple 
region) for $\abprime >0$ and for all values of $\tanb$ (when compared to the same 
obtained for the $M_h^{125}$-MSSM scenario) and a reinforced one (to the edge of the 
pink region) for $\abprime <0$ only when $\tanb$ is on the larger side
($\geqslant 30$). Note that for the curves showing maximal deviations in
$\sigma \times \mathrm{BR}$ in figure \ref{subfig:sigbr-tb-abp} and hence maximal 
alterations in the exclusion limits in figure \ref{subfig:ma-tb-excl}, with respect 
to the $\abprime=0$ case, are obtained with combinations of $\abtauprime$ with 
maximal values we use for them (i.e., $|\abtauprime|=3$ TeV) and setting 
$\mathrm{sign}(\atauprime)$ to be the same as that on $\abprime$ which broadly controls the nature of such alterations. We have checked that all these findings also agree closely with the exclusion obtained using {\tt HiggsBounds-v5.9.1} which now incorporates the pertinent ATLAS analysis carried out in reference \cite{ATLAS:2020zms}.

The nature of the modified exclusion contours could be broadly reconciled by 
referring to the same for the variations of BR($H/A \to \tautaubar$) and the 
cross section $\sigma_{(pp \to \bbbar H/A)}$ as illustrated in figures 
\ref{fig:br-tautau-abp-ataup} and \ref{fig:xsec-ma-tanb}, respectively. Of 
particular interest is the asymmetric nature of the possible relaxations and 
reinforcements in the exclusions in the NHSSM scenario that are obtained using 
somewhat large (extremal) values of $\abprime$ and $\tanb$, when compared to 
the exclusions obtained for the MSSM case. Note from figure
\ref{fig:br-tautau-abp-ataup} that BR($H/A \to \tautaubar$) more or less 
saturates as $\abprime$ and $\tanb$ approach large values irrespective of the 
sign on $\abprime$. However, while for large, positive $\abprime$,
BR($H/A \to \tautaubar$) reaches a plateau at a larger value compared to the MSSM case, the reverse is true for large, negative $\abprime$. On top of that, from figures
\ref{subfig:xsec-ma-tanb-abp3} and \ref{subfig:xsec-ma-tanb-abp0} (the MSSM case) 
one can find that the cross sections in these two scenarios for any given
$\tanb$ and $\abprime$ are not too different. Hence the product
$\sigma \times \mathrm{BR}$ remains nearly as sensitive to the experiment as 
for the MSSM case. This is behind a minor relaxation in $m_A$ ($\lesssim 100$ GeV) in the exclusion even though the setup is tailored to bring about the maximum possible exclusion. On the other hand, it can be gleaned from figure \ref{subfig:xsec-ma-tanb-abp-3} that for large, negative $\abprime$,
the difference in the cross sections $\sigma_{(pp \to \bbbar H/A)}$ with that from the MSSM scenario quickly grows for growing $m_A$ and $\tanb$. Though BR($H/A \to \tautaubar$) tends
to saturate at a value lower than that in the MSSM case, it is entirely because of the rapid growth in the
cross section with $\tanb$ that drives the sensitivity to the experiment and hence a stronger exclusion of the $m_A-\tanb$ plane (by $\sim 200$ GeV in $m_A$, for $\tanb=60$) for large, negative
$\abprime$. At a more fundamental level, these features are perhaps best understood in terms of the variations of
$\ybtau(\tanb,\abprime)$ as shown in figure \ref{subfig:yf-tb-afp} and then, of $g^2_{Hbb}(\tanb,\abprime)$ as shown in figures \ref{subfig:gsq-bbphi-tb} and \ref{subfig:gsq-tautauphi-tb}.

In this context, it is noteworthy that the extents of relaxation and reinforcement of exclusions that are presented in figure
\ref{subfig:ma-tb-excl} both refer to NHSSM scenarios that are cut out for
yielding such alterations to be nearly maximal. The only way such modifications could get somewhat pronounced is by setting $\mgluino$ at a value higher than what we have used, i.e., 3 TeV (in fact, this is the
smallest value from the range we have used to draw the plots in figure
\ref{fig:msbot-gsq-abp}. On these plots, at $|\abprime|=\msb$, the widths of the bands are the largest and we are referring to the lowest (highest) points from
these bands at $|\abprime|=\msb$ for $\abprime <0 \, (>0)$.     
%
\section{Conclusions}
\label{sec:conclude}
In this work, we have discussed in detail the nature of the dependencies of each 
of the bottom and the tau Yukawa couplings simultaneously on the NHSSM-specific 
trilinear soft parameters $\abprime$ and $\atauprime$, respectively, and on
$\tanb$. The MSSM higgsino mass parameter is held at not too large a value (500 
GeV) thus aiding the scenario to remain somewhat `natural'. We have further 
undertaken a thorough investigation on how these could affect, in the NHSSM 
scenario, the production cross sections of the heavy neutral Higgs bosons ($H$ 
and $A$) in association with a $b$-quark pair and their branching fractions to
$\tautaubar$, the two key ingredients that dictate their reach at the current 
and future runs of the LHC. These are demonstrated in terms of altered (relaxed 
or extended) exclusion regions in the customary $m_A-\tanb$ plane and 
contrasted with the latest such constraint from the LHC for an MSSM scenario, 
for some representative values of $\abprime$. The role of $\atauprime$ is found 
to be rather limited while the same of the masses of the sbottom and the 
gluino can be moderate.

A general finding is that such deviations are maximal when both $\tanb$ and the 
magnitude of $\abprime$ are large.  For an example, for $\tanb=60$, while a 
maximum relaxation of $\approx 100$ GeV in $m_{A}$ is observed in the NHSSM 
scenario for a large positive $\abprime$ ($=3$ TeV) when in the MSSM it is 
excluded up to 2 TeV, a strengthening of the exclusion by about 200 GeV with 
respect to the latter occurs for a large negative $\abprime$ ($=-3$ TeV). For smaller values of $\tanb$, the role of $\abprime$ is set to diminish since, in that regime, $\abprime$ affects $\sigma(pp \to \bbbar H,A)$ and
BR[$H,A \to \tautaubar$] in a complementary way such that the all-important quantity $\sigma \times \mathrm{BR}$ ceases to be sensitive to a variation in
$\abprime$. Thus, for smaller values of $\tanb$, the exclusion contour in the $m_A-\tanb$ plane in the NHSSM scenario does not deviate much from the reported (MSSM) one. Overall, unless for a rather large value of $\tanb$, the relevant NHSSM parameters tend to conspire such that the LHC sensitivity of the scenario does not differ much from the MSSM case. As far as the exclusion of the $m_A-\tanb$ parameter plane is concerned, this pattern is expected to be broadly preserved at the upcoming runs of the LHC.
%


%
\section*{Acknowledgements}
AKS thanks the Department of Physics, SGTB Khalsa College, University of Delhi and the SERB sponsored Multi Institutional Project titled ``Probing New Physics Interactions'' (CRG/2018/004889) running there under which the major part of the work has been carried out.
The authors also like to thank Mark Goodsell for many useful discussions on various aspects of the implementation of the NHSSM scenario in {\tt SPheno} via {\tt SARAH}.
They also thank Arnaud Ferrari from the ATLAS Collaboration, Rikkert Frederix, Jean-Loic Kneur, Davide Napoletano, Werner Porod, V. Ravindran  and Michael Spira for very helpful exchanges.

%

%

\begin{thebibliography}{999}
%
\bibitem{Aad:2019mbh}
G.~Aad \textit{et al.} [ATLAS],
Phys. Rev. D \textbf{101} (2020) no.1, 012002
doi:10.1103/PhysRevD.101.012002
[arXiv:1909.02845 [hep-ex]].

\bibitem{Wiesemann:2014ioa}
M.~Wiesemann, R.~Frederix, S.~Frixione, V.~Hirschi, F.~Maltoni and P.~Torrielli,
JHEP \textbf{02} (2015), 132
doi:10.1007/JHEP02(2015)132
[arXiv:1409.5301 [hep-ph]].

\bibitem{Pagani:2020rsg}
D.~Pagani, H.~S.~Shao and M.~Zaro,
JHEP \textbf{11} (2020), 036
doi:10.1007/JHEP11(2020)036
[arXiv:2005.10277 [hep-ph]].

\bibitem{Sirunyan:2018zut}
A.~M.~Sirunyan \textit{et al.} [CMS],
JHEP \textbf{09} (2018), 007
doi:10.1007/JHEP09(2018)007
[arXiv:1803.06553 [hep-ex]].

\bibitem{Aad:2020zxo}
G.~Aad \textit{et al.} [ATLAS],
Phys. Rev. Lett. \textbf{125} (2020) no.5, 051801
doi:10.1103/PhysRevLett.125.051801
[arXiv:2002.12223 [hep-ex]].

\bibitem{Aaboud:2017sjh}
M.~Aaboud \textit{et al.} [ATLAS],
JHEP \textbf{01} (2018), 055
doi:10.1007/JHEP01(2018)055
[arXiv:1709.07242 [hep-ex]].

\bibitem{Girardello:1981wz}
L.~Girardello and M.~T.~Grisaru,
Nucl. Phys. B \textbf{194} (1982), 65
doi:10.1016/0550-3213(82)90512-0

\bibitem{Martin:1999hc}
S.~P.~Martin,
Phys. Rev. D \textbf{61} (2000), 035004
doi:10.1103/PhysRevD.61.035004
[arXiv:hep-ph/9907550 [hep-ph]].


\bibitem{Ross:2016pml}
G.~G.~Ross, K.~Schmidt-Hoberg and F.~Staub,
Phys.\ Lett.\ B {\bf 759} (2016) 110
doi:10.1016/j.physletb.2016.05.053
[arXiv:1603.09347 [hep-ph]].

\bibitem{Ross:2017kjc}
G.~G.~Ross, K.~Schmidt-Hoberg and F.~Staub,
JHEP {\bf 1703} (2017) 021
doi:10.1007/JHEP03(2017)021
[arXiv:1701.03480 [hep-ph]].


\bibitem{Chattopadhyay:2017qvh}
U.~Chattopadhyay, D.~Das and S.~Mukherjee,
JHEP \textbf{01} (2018), 158
doi:10.1007/JHEP01(2018)158
[arXiv:1710.10120 [hep-ph]].

\bibitem{Chakraborty:2019wav}
S.~Chakraborty and T.~S.~Roy,
Phys. Rev. D \textbf{100}, no.3, 035020 (2019)
doi:10.1103/PhysRevD.100.035020
[arXiv:1904.10144 [hep-ph]].

\bibitem{Haber:2007dj}
H.~E.~Haber and J.~D.~Mason,
Phys. Rev. D \textbf{77} (2008), 115011
doi:10.1103/PhysRevD.77.115011
[arXiv:0711.2890 [hep-ph]].

\bibitem{Bagger:1993ji}
J.~Bagger and E.~Poppitz,
Phys. Rev. Lett. \textbf{71} (1993), 2380-2382
doi:10.1103/PhysRevLett.71.2380
[arXiv:hep-ph/9307317 [hep-ph]].

\bibitem{Ellwanger:1983mg}
U.~Ellwanger,
Phys. Lett. B \textbf{133} (1983), 187-191
doi:10.1016/0370-2693(83)90557-9

\bibitem{Jack:1999ud}
I.~Jack and D.~R.~T.~Jones,
Phys. Lett. B \textbf{457} (1999), 101-108
doi:10.1016/S0370-2693(99)00530-4
[arXiv:hep-ph/9903365 [hep-ph]].

\bibitem{Chattopadhyay:2016ivr}
U.~Chattopadhyay and A.~Dey,
JHEP \textbf{10} (2016), 027
doi:10.1007/JHEP10(2016)027
[arXiv:1604.06367 [hep-ph]].



\bibitem{Chattopadhyay:2018tqv}
U.~Chattopadhyay, A.~Datta, S.~Mukherjee and A.~K.~Swain,
JHEP \textbf{10} (2018), 202
doi:10.1007/JHEP10(2018)202
[arXiv:1809.05438 [hep-ph]].

\bibitem{Hall:1993gn}
L.~J.~Hall, R.~Rattazzi and U.~Sarid,
Phys. Rev. D \textbf{50} (1994), 7048-7065
doi:10.1103/PhysRevD.50.7048
[arXiv:hep-ph/9306309 [hep-ph]].

\bibitem{Hempfling:1993kv}
R.~Hempfling,
Phys. Rev. D \textbf{49} (1994), 6168-6172
doi:10.1103/PhysRevD.49.6168

\bibitem{Carena:1994bv}
M.~Carena, M.~Olechowski, S.~Pokorski and C.~E.~M.~Wagner,
Nucl. Phys. B \textbf{426} (1994), 269-300
doi:10.1016/0550-3213(94)90313-1
[arXiv:hep-ph/9402253 [hep-ph]].

\bibitem{Pierce:1996zz}
D.~M.~Pierce, J.~A.~Bagger, K.~T.~Matchev and R.~j.~Zhang,
Nucl. Phys. B \textbf{491} (1997), 3-67
doi:10.1016/S0550-3213(96)00683-9
[arXiv:hep-ph/9606211 [hep-ph]].

\bibitem{Logan:2000cz}
H.~E.~Logan,
Nucl. Phys. B Proc. Suppl. \textbf{101} (2001), 279-288
doi:10.1016/S0920-5632(01)01512-2
[arXiv:hep-ph/0102029 [hep-ph]].

\bibitem{Antusch:2008tf}
S.~Antusch and M.~Spinrath,
Phys. Rev. D \textbf{78} (2008), 075020
doi:10.1103/PhysRevD.78.075020
[arXiv:0804.0717 [hep-ph]].

\bibitem{Carena:2002es}
M.~Carena and H.~E.~Haber,
Prog. Part. Nucl. Phys. \textbf{50}, 63-152 (2003)
doi:10.1016/S0146-6410(02)00177-1
[arXiv:hep-ph/0208209 [hep-ph]].


\bibitem{Djouadi:2005gj}
A.~Djouadi,
Phys. Rept. \textbf{459} (2008), 1-241
doi:10.1016/j.physrep.2007.10.005
[arXiv:hep-ph/0503173 [hep-ph]].


\bibitem{ATLAS:2018gfm}
M.~Aaboud \textit{et al.} [ATLAS],
JHEP \textbf{09} (2018), 139
doi:10.1007/JHEP09(2018)139
[arXiv:1807.07915 [hep-ex]].

\bibitem{ATLAS:2021upq}
G.~Aad \textit{et al.} [ATLAS],
JHEP \textbf{06} (2021), 145
doi:10.1007/JHEP06(2021)145
[arXiv:2102.10076 [hep-ex]].

\bibitem{ATLAS:2021yij}
G.~Aad \textit{et al.} [ATLAS],
JHEP \textbf{05} (2021), 093
doi:10.1007/JHEP05(2021)093
[arXiv:2101.12527 [hep-ex]].

\bibitem{Dedes:1998yt}
A.~Dedes and S.~Moretti,
Phys. Rev. D \textbf{60} (1999), 015007
doi:10.1103/PhysRevD.60.015007
[arXiv:hep-ph/9812328 [hep-ph]].

\bibitem{Beuria:2017gtf}
J.~Beuria and A.~Dey,
JHEP \textbf{10} (2017), 154
doi:10.1007/JHEP10(2017)154
[arXiv:1708.08361 [hep-ph]].



\bibitem{Carena:1999py}
M.~Carena, D.~Garcia, U.~Nierste and C.~E.~M.~Wagner,
Nucl. Phys. B \textbf{577}, 88-120 (2000)
doi:10.1016/S0550-3213(00)00146-2
[arXiv:hep-ph/9912516 [hep-ph]].


\bibitem{Guasch:2003cv}
J.~Guasch, P.~Hafliger and M.~Spira,
Phys. Rev. D \textbf{68}, 115001 (2003)
doi:10.1103/PhysRevD.68.115001
[arXiv:hep-ph/0305101 [hep-ph]].


\bibitem{Noth:2008tw}
D.~Noth and M.~Spira,
Phys. Rev. Lett. \textbf{101}, 181801 (2008)
doi:10.1103/PhysRevLett.101.181801
[arXiv:0808.0087 [hep-ph]].


\bibitem{Noth:2010jy}
D.~Noth and M.~Spira,
JHEP \textbf{06}, 084 (2011)
doi:10.1007/JHEP06(2011)084
[arXiv:1001.1935 [hep-ph]].


\bibitem{Ghezzi:2017enb}
M.~Ghezzi, S.~Glaus, D.~M\"uller, T.~Schmidt and M.~Spira,
Eur. Phys. J. C \textbf{81}, no.3, 259 (2021)
doi:10.1140/epjc/s10052-021-09035-6
[arXiv:1711.02555 [hep-ph]].


\bibitem{Mihaila:2010mp}
L.~Mihaila and C.~Reisser,
JHEP \textbf{08}, 021 (2010)
doi:10.1007/JHEP08(2010)021
[arXiv:1007.0693 [hep-ph]].

\bibitem{Crivellin:2010er}
A.~Crivellin,
Phys. Rev. D \textbf{83} (2011), 056001
doi:10.1103/PhysRevD.83.056001
[arXiv:1012.4840 [hep-ph]].

\bibitem{Crivellin:2011jt}
A.~Crivellin, L.~Hofer and J.~Rosiek,
JHEP \textbf{07} (2011), 017
doi:10.1007/JHEP07(2011)017
[arXiv:1103.4272 [hep-ph]].

\bibitem{Crivellin:2012zz}
A.~Crivellin and C.~Greub,
Phys. Rev. D \textbf{87} (2013), 015013
[erratum: Phys. Rev. D \textbf{87} (2013), 079901]
doi:10.1103/PhysRevD.87.015013
[arXiv:1210.7453 [hep-ph]].
%
\bibitem{Girrbach:2009uy}
J.~Girrbach, S.~Mertens, U.~Nierste and S.~Wiesenfeldt,
JHEP \textbf{05} (2010), 026
doi:10.1007/JHEP05(2010)026
[arXiv:0910.2663 [hep-ph]].
%
%

\bibitem{Gunion:1984yn}
J.~F.~Gunion and H.~E.~Haber,
Nucl. Phys. B \textbf{272} (1986), 1
[erratum: Nucl. Phys. B \textbf{402} (1993), 567-569]
doi:10.1016/0550-3213(86)90340-8
%

\bibitem{Gunion:1989we}
J.~F.~Gunion, H.~E.~Haber, G.~L.~Kane and S.~Dawson,
Front. Phys. \textbf{80} (2000), 1-404
SCIPP-89/13.


\bibitem{Dawson:2005vi}
S.~Dawson, C.~B.~Jackson, L.~Reina and D.~Wackeroth,
Mod. Phys. Lett. A \textbf{21} (2006), 89-110
doi:10.1142/S0217732306019256
[arXiv:hep-ph/0508293 [hep-ph]].


\bibitem{Staub:2013tta}
F.~Staub,
Comput. Phys. Commun. \textbf{185} (2014), 1773-1790
doi:10.1016/j.cpc.2014.02.018
[arXiv:1309.7223 [hep-ph]].

\bibitem{Staub:2015kfa}
F.~Staub,
Adv. High Energy Phys. \textbf{2015} (2015), 840780
doi:10.1155/2015/840780
[arXiv:1503.04200 [hep-ph]].

\bibitem{Porod:2011nf}
W.~Porod and F.~Staub,
Comput. Phys. Commun. \textbf{183} (2012), 2458-2469
doi:10.1016/j.cpc.2012.05.021
[arXiv:1104.1573 [hep-ph]].

\bibitem{Bechtle:2020pkv}
P.~Bechtle, D.~Dercks, S.~Heinemeyer, T.~Klingl, T.~Stefaniak, G.~Weiglein and J.~Wittbrodt,
Eur. Phys. J. C \textbf{80} (2020) no.12, 1211
doi:10.1140/epjc/s10052-020-08557-9
[arXiv:2006.06007 [hep-ph]].

\bibitem{Campbell:2004pu}
J.~M.~Campbell, S.~Dawson, S.~Dittmaier, C.~Jackson, M.~Kramer, F.~Maltoni, L.~Reina, M.~Spira, D.~Wackeroth and S.~Willenbrock,
[arXiv:hep-ph/0405302 [hep-ph]].


\bibitem{Barnett:1987jw}
R.~M.~Barnett, H.~E.~Haber and D.~E.~Soper,
Nucl. Phys. B \textbf{306} (1988), 697-745
doi:10.1016/0550-3213(88)90440-3

\bibitem{Dicus:1988cx}
D.~A.~Dicus and S.~Willenbrock,
Phys. Rev. D \textbf{39} (1989), 751
doi:10.1103/PhysRevD.39.751

\bibitem{Harlander:2011aa}
R.~Harlander, M.~Kramer and M.~Schumacher,
[arXiv:1112.3478 [hep-ph]].

\bibitem{ATLAS:2017eiz}
M.~Aaboud \textit{et al.} [ATLAS],
JHEP \textbf{01} (2018), 055
doi:10.1007/JHEP01(2018)055
[arXiv:1709.07242 [hep-ex]].

\bibitem{CMS:2018rmh}
A.~M.~Sirunyan \textit{et al.} [CMS],
JHEP \textbf{09} (2018), 007
doi:10.1007/JHEP09(2018)007
[arXiv:1803.06553 [hep-ex]].

\bibitem{ATLAS:2020zms}
G.~Aad \textit{et al.} [ATLAS],
Phys. Rev. Lett. \textbf{125} (2020) no.5, 051801
doi:10.1103/PhysRevLett.125.051801
[arXiv:2002.12223 [hep-ex]].

\bibitem{Dicus:1998hs}
D.~Dicus, T.~Stelzer, Z.~Sullivan and S.~Willenbrock,
Phys. Rev. D \textbf{59} (1999), 094016
doi:10.1103/PhysRevD.59.094016
[arXiv:hep-ph/9811492 [hep-ph]].

\bibitem{Balazs:1998sb}
C.~Balazs, H.~J.~He and C.~P.~Yuan,
Phys. Rev. D \textbf{60} (1999), 114001
doi:10.1103/PhysRevD.60.114001
[arXiv:hep-ph/9812263 [hep-ph]].

\bibitem{Harlander:2003ai}
R.~V.~Harlander and W.~B.~Kilgore,
Phys. Rev. D \textbf{68} (2003), 013001
doi:10.1103/PhysRevD.68.013001
[arXiv:hep-ph/0304035 [hep-ph]].

\bibitem{Dittmaier:2003ej}
S.~Dittmaier, M.~Kr\"amer and M.~Spira,
Phys. Rev. D \textbf{70} (2004), 074010
doi:10.1103/PhysRevD.70.074010
[arXiv:hep-ph/0309204 [hep-ph]].


\bibitem{Dawson:2003kb}
S.~Dawson, C.~B.~Jackson, L.~Reina and D.~Wackeroth,
Phys. Rev. D \textbf{69} (2004), 074027
doi:10.1103/PhysRevD.69.074027
[arXiv:hep-ph/0311067 [hep-ph]].




\bibitem{Duhr:2019kwi}
C.~Duhr, F.~Dulat and B.~Mistlberger,
Phys. Rev. Lett. \textbf{125} (2020) no.5, 051804
doi:10.1103/PhysRevLett.125.051804
[arXiv:1904.09990 [hep-ph]].

\bibitem{Ajjath:2019neu}
A.~H.~Ajjath, A.~Chakraborty, G.~Das, P.~Mukherjee and V.~Ravindran,
JHEP \textbf{11} (2019), 006
doi:10.1007/JHEP11(2019)006
[arXiv:1905.03771 [hep-ph]].

\bibitem{Aivazis:1993pi}
M.~A.~G.~Aivazis, J.~C.~Collins, F.~I.~Olness and W.~K.~Tung,
Phys. Rev. D \textbf{50} (1994), 3102-3118
doi:10.1103/PhysRevD.50.3102
[arXiv:hep-ph/9312319 [hep-ph]].

\bibitem{Thorne:1997ga}
R.~S.~Thorne and R.~G.~Roberts,
Phys. Rev. D \textbf{57} (1998), 6871-6898
doi:10.1103/PhysRevD.57.6871
[arXiv:hep-ph/9709442 [hep-ph]].

\bibitem{Cacciari:1998it}
M.~Cacciari, M.~Greco and P.~Nason,
JHEP \textbf{05} (1998), 007
doi:10.1088/1126-6708/1998/05/007
[arXiv:hep-ph/9803400 [hep-ph]].

\bibitem{Kramer:2000hn}
M.~Kr\"amer, F.~I.~Olness and D.~E.~Soper,
Phys. Rev. D \textbf{62} (2000), 096007
doi:10.1103/PhysRevD.62.096007
[arXiv:hep-ph/0003035 [hep-ph]].

\bibitem{Tung:2001mv}
W.~K.~Tung, S.~Kretzer and C.~Schmidt,
J. Phys. G \textbf{28} (2002), 983-996
doi:10.1088/0954-3899/28/5/321
[arXiv:hep-ph/0110247 [hep-ph]].

\bibitem{Thorne:2006qt}
R.~S.~Thorne,
Phys. Rev. D \textbf{73} (2006), 054019
doi:10.1103/PhysRevD.73.054019
[arXiv:hep-ph/0601245 [hep-ph]].

\bibitem{Forte:2015hba}
S.~Forte, D.~Napoletano and M.~Ubiali,
Phys. Lett. B \textbf{751} (2015), 331-337
doi:10.1016/j.physletb.2015.10.051
[arXiv:1508.01529 [hep-ph]].

\bibitem{Bonvini:2015pxa}
M.~Bonvini, A.~S.~Papanastasiou and F.~J.~Tackmann,
JHEP \textbf{11} (2015), 196
doi:10.1007/JHEP11(2015)196
[arXiv:1508.03288 [hep-ph]].

\bibitem{Bonvini:2016fgf}
M.~Bonvini, A.~S.~Papanastasiou and F.~J.~Tackmann,
JHEP \textbf{10} (2016), 053
doi:10.1007/JHEP10(2016)053
[arXiv:1605.01733 [hep-ph]].

\bibitem{Forte:2016sja}
S.~Forte, D.~Napoletano and M.~Ubiali,
Phys. Lett. B \textbf{763} (2016), 190-196
doi:10.1016/j.physletb.2016.10.040
[arXiv:1607.00389 [hep-ph]].

\bibitem{Duhr:2020kzd}
C.~Duhr, F.~Dulat, V.~Hirschi and B.~Mistlberger,
JHEP \textbf{08} (2020) no.08, 017
doi:10.1007/JHEP08(2020)017
[arXiv:2004.04752 [hep-ph]].

\bibitem{LHCHiggsCrossSectionWorkingGroup:2016ypw}
D.~de Florian \textit{et al.} [LHC Higgs Cross Section Working Group],
doi:10.23731/CYRM-2017-002
[arXiv:1610.07922 [hep-ph]].

\bibitem{recommend}
See {\footnotesize \tt ``https://twiki.cern.ch/twiki/bin/view/LHCPhysics/LHCHWGBBH\#NLO\_NNLLpart\_ybyt\_matching''}.

\bibitem{Maltoni:2003pn}
F.~Maltoni, Z.~Sullivan and S.~Willenbrock,
Phys. Rev. D \textbf{67} (2003), 093005
doi:10.1103/PhysRevD.67.093005
[arXiv:hep-ph/0301033 [hep-ph]].

\bibitem{Boos:2003yi}
E.~Boos and T.~Plehn,
Phys. Rev. D \textbf{69} (2004), 094005
doi:10.1103/PhysRevD.69.094005
[arXiv:hep-ph/0304034 [hep-ph]].

\bibitem{Maltoni:2005wd}
F.~Maltoni, T.~McElmurry and S.~Willenbrock,
Phys. Rev. D \textbf{72} (2005), 074024
doi:10.1103/PhysRevD.72.074024
[arXiv:hep-ph/0505014 [hep-ph]].

\bibitem{Maltoni:2012pa}
F.~Maltoni, G.~Ridolfi and M.~Ubiali,
JHEP \textbf{07} (2012), 022
[erratum: JHEP \textbf{04} (2013), 095]
doi:10.1007/JHEP04(2013)095
[arXiv:1203.6393 [hep-ph]].

\bibitem{Harlander:2015xur}
R.~V.~Harlander,
Eur. Phys. J. C \textbf{76} (2016) no.5, 252
doi:10.1140/epjc/s10052-016-4093-x
[arXiv:1512.04901 [hep-ph]].

\bibitem{Dittmaier:2006cz}
S.~Dittmaier, M.~Kr\"amer, A.~Muck and T.~Schluter,
JHEP \textbf{03} (2007), 114
doi:10.1088/1126-6708/2007/03/114
[arXiv:hep-ph/0611353 [hep-ph]].

\bibitem{Dawson:2011pe}
S.~Dawson, C.~B.~Jackson and P.~Jaiswal,
Phys. Rev. D \textbf{83} (2011), 115007
doi:10.1103/PhysRevD.83.115007
[arXiv:1104.1631 [hep-ph]].

\bibitem{Dittmaier:2014sva}
S.~Dittmaier, P.~H\"afliger, M.~Kr\"amer, M.~Spira and M.~Walser,
Phys. Rev. D \textbf{90} (2014) no.3, 035010
doi:10.1103/PhysRevD.90.035010
[arXiv:1406.5307 [hep-ph]].

\bibitem{Bagnaschi:2018ofa}
E.~Bagnaschi, H.~Bahl, E.~Fuchs, T.~Hahn, S.~Heinemeyer, S.~Liebler, S.~Patel, P.~Slavich, T.~Stefaniak and C.~E.~M.~Wagner, \textit{et al.}
Eur. Phys. J. C \textbf{79} (2019) no.7, 617
doi:10.1140/epjc/s10052-019-7114-8
[arXiv:1808.07542 [hep-ph]].
%
\end{thebibliography}
\end{document}